\documentclass[apj, numberedappendix, appendixfloats]{emulateapj}
\usepackage{natbib}
\usepackage{amsmath}
\usepackage{enumerate}
\usepackage{hyperref}

\received{}
\revised{}
\accepted{\today}

\shorttitle{OLAS I: Gas Kinematics in Dwarf Galaxies}
\shortauthors{Hirtenstein et al.}

\begin{document}

\title{The OSIRIS Lens-Amplified Survey (OLAS) I:
Dynamical effects of stellar feedback in low mass galaxies at z $\sim$ 2}

\email{jhirtenstein@ucdavis.edu}

\author{Jessie Hirtenstein\altaffilmark{1}}
\author{Tucker Jones\altaffilmark{1}}
\author{Xin Wang\altaffilmark{2}}
\author{Andrew Wetzel\altaffilmark{1}}
\author{Kareem El-Badry\altaffilmark{3}}
\author{Austin Hoag\altaffilmark{2}}
\author{Tommaso Treu\altaffilmark{2}}
\author{Maru\v{s}a Brada\v{c}\altaffilmark{1}}
\author{Takahiro Morishita\altaffilmark{4}}
\affil{\altaffilmark{1}Department of Physics, University of California, Davis, CA, USA}
\affil{\altaffilmark{2}Department of Physics and Astronomy, University of California, Los Angeles, CA, USA}
\affil{\altaffilmark{3}Department of Astronomy, University of California, Berkeley, CA, USA}
\affil{\altaffilmark{4}Space Telescope Science Institute, 3700 San Martin Drive, Baltimore, MD 21218, USA}





\begin{abstract}
We introduce the OSIRIS Lens-Amplified Survey (OLAS), a kinematic survey of gravitationally lensed galaxies at cosmic noon taken with Keck adaptive optics. In this paper we present spatially resolved spectroscopy and nebular emission kinematic maps for 17 star forming galaxies with stellar masses $8<$ log$(M_*/M_{\odot})<9.8$ and redshifts $1.2<z<2.3$. OLAS is designed to probe the stellar mass (M$_*$) and specific star formation rate (sSFR) range where simulations suggest that stellar feedback is most effective at driving gaseous outflows that create galaxy-wide potential fluctuations which can generate dark matter cores. We compare our kinematic data with the trend between sSFR, M$_*$ and H$\alpha$ velocity dispersion, $\sigma$, from the Feedback In Realistic Environments (FIRE) simulations. Our observations reveal a correlation between sSFR and $\sigma$ at fixed M$_*$ that is similar to the trend predicted by simulations: feedback from star formation drives star-forming gas and newly formed stars into more dispersion dominated orbits. The observed magnitude of this effect is in good agreement with the FIRE simulations, in which feedback alters the central density profiles of low mass galaxies, converting dark matter cusps into cores over time. Our data support the scenario that stellar feedback drives gaseous outflows and potential fluctuations, which in turn drive dark matter core formation in dwarf galaxies.

\end{abstract}




\section{Introduction} \label{sec:intro}

The standard cosmological model has been tremendously successful in predicting the large-scale structure and distribution of galaxies in terms of dark energy and cold dark matter ($\Lambda$CDM; e.g. \citealt{Planck_2018}). However in the most common systems, dwarf galaxies (log$(M_*/M_{\odot}$) $\lesssim$ 8 - 9), there is tension between the predictions of dark-matter only simulations and observations. Dark-matter only models predict steep central dark matter profiles (referred to as cusps; \citealt{Navarro_1996}; \citealt{Navarro_2010}), yet observed dwarf galaxies in the local universe reveal dark matter halos with a diversity of profiles, with many having nearly constant-density cores (e.g. \citealt{McGaugh_2001}; \citealt{deBlok_2008}). This ``cusp-core" problem is one of the most significant and long-standing challenges to standard $\Lambda$CDM and has motivated alternative dark matter theories such as warm dark matter (e.g. \citealt{Spergel_2000}; \citealt{Viel_2005}) and self-interacting dark matter (e.g. \citealt{Buckley_Fox_2010}; \citealt{Zavala_2013}; \citealt{Vogelsberger_2014}) to resolve this tension, among others. Thus, understanding the origin of the cusp-core problem is vital to understanding the dominant mass component of the universe.

One exciting resolution to the cusp-core problem that does not require abandoning $\Lambda$CDM is to include baryonic physics in cosmological simulations. Recent works show stellar feedback may be capable of altering the dark matter distribution of dwarf galaxies within the local group (e.g. \citealt{Governato_2012}; \citealt{Pontzen_2012}; \citealt{Chan_2015}; \citealt{Onorbe_2015}; \citealt{Read_2016}, \citealt{ElBadry_2016}; \citealt{ElBadry_2017}; \citealt{Read_2018}). These studies show that strong baryonic outflows at sub-kpc scales significantly lower the central density of dark matter halos, resolving the cusp-core problem and revealing the mass range where feedback most significantly drives core formation: stellar masses M$_* \sim 10^6$ - $10^9 M_{\odot}$. 

However, not all stellar feedback models alter the central density profile of dwarf galaxies (i.e., \citealt{Bose_2018}). Further, the models that successfully resolve the cusp-core problem in local group dwarfs have different expectations at high redshifts (e.g., \citealt{Gibson_2013}), such as competing implications for galaxy masses, chemical enrichment and pollution of the interstellar and intergalactic media (e.g., \citealt{Dave_2011}).

\begin{table*}
\centering
\caption{Overview of observational data}
\resizebox{0.9\textwidth}{!}{%
	\begin{tabular}{c c c c c c c c c c c c c c}
		\hline
		Cluster & ID &  z  & RA & Dec &  Dates  & AO &  Filter  & Scale & $t_{exp}$  & PSF\footnote{Native point spread functions (PSFs) derived from the full width half max of the tip/tilt star} & $\mu$ & $m_{F160}$\footnote{AB magnitude measured in the HST/WFC3-IR F160W filter} & \\
		 \ & \ &  \ & \ &  \ & MM/DD/YY & \ & \ & $''$ & $s$ & $''$ & \ &\\
		\hline
		A370 & 02056 & 1.27 & 02:39:50.260 & -01:34:24.20 & 11/01/17 & LGS & Hn1 & 0.05 & 3000 & 0.096 & $1.90^{+0.04}_{-0.02}$ & 23.1\\
		A370 & 03097 & 1.55 & 02:39:50.270 & -01:35:02.70 & 10/30/17 & NGS & Hn4 & 0.1 & 10800 & 0.165 & $2.30^{+0.11}_{-0.10}$ & 22.4\\
		A370 & 03312\footnote{This object was determined to be an AGN and is excluded from this analysis} &1.60 & 02:39:47.543 & -01:35:12.23 & 10/30/17 & NGS & Hn4 & 0.1 & 5400 & 0.135 & $1.90^{+0.04}_{-0.04}$ & 22.5\\
		M0717 & 01828 & 1.47 & 07:17:39.367 & +37:44:31.88 & 10/25-26/18 & LGS & Hn3 & 0.05 & 6600 & 0.131 & $3.02^{+0.15}_{-0.14}$ & 22.9\\
		M0717 & 02064 & 2.07 & 07:17:39.125 & +37:44:18.45 & 11/01/17 & LGS & Kn1 & 0.1 & 7200 & 0.136 & $6.48^{+0.97}_{-0.79}$ & 23.3\\
		M0744 & 00920 & 1.28 & 07:44:50.950 & +39:27:35.80 & 10/29/17 & NGS & Hn1 & 0.1 & 7200 & 0.192 & $20.2^{+2.29}_{-1.72}$ & 20.7\\
		M0744 & 01203 & 1.65 & 07:44:47.420 & +39:27:24.10 & 10/21/16 & LGS & Hn5 & 0.05 & 2700 & 0.092 & $3.16^{+0.06}_{-0.06}$ & 22.0\\
		M0744 & 02341 & 1.28 & 07:44:51.053 & +39:26:27.82 & 10/24/18 & LGS & Hn1 & 0.05 & 4500 & 0.281 & $2.09^{+0.16}_{-0.07}$ & 22.5\\
		M1149 & 00593 & 1.48 & 11:49:37.661 & +22:24:27.00 & 06/13/17 & LGS & Hn3 & 0.1 & 3000 & 0.281 & $1.54^{+0.01}_{-0.00}$ & 22.2\\
		M1149 & 00683 & 1.68 & 11:49:35.294 & +22:24:22.28 & 06/12/17 & LGS & Hn5 & 0.1 & 3600 & 0.149 & $4.05^{+0.02}_{-0.01}$ & 25.5\\
		M1149 & 01058 & 1.25 & 11:49:34.044 & +22:24:00.38 & 06/12/17 & LGS & Hn1 & 0.05 & 4500 & 0.104 & $2.52^{+0.01}_{-0.00}$ & 22.5\\
		M1149 & 01802 & 2.16 & 11:49:39.358 & +22:23:09.06 & 05/31/18 & LGS & Kn2 & 0.05 & 7200 & 0.062 & $2.42^{+0.03}_{-0.03}$ & 23.1\\
		M1423 & 00248 & 1.42 & 14:23:50.124 & +24:05:32.17 & 06/12/17 & LGS & Hn2 & 0.1 & 4320 & 0.164 & $2.02^{+0.07}_{-0.07}$ & 21.7\\
		M2129 & 00465 & 1.36 & 21:29:28.174 & -07:40:54.71 & 10/25/18 & LGS & Hn2 & 0.05 & 4500 & 0.127 & $1.74^{+0.08}_{-0.04}$ & 22.1\\
		M2129 & 00478 & 1.67 & 21:29:24.511 & -07:40:54.79 & 06/12/17 & LGS & Hn5 & 0.1 & 1800 & 0.228 & $1.79^{+0.04}_{-0.03}$ & 24.5\\
		M2129 & 01408 & 1.48 & 21:29:28.569 & -07:42:01.95 & 06/13/17 & LGS & Hn3 & 0.05 & 5400 & 0.207 & $1.86^{+0.17}_{-0.13}$ & 23.5\\
		M2129 & 01665 & 1.56 & 21:29:25.956 & -07:42:24.15 & 10/25/18 & LGS & Hn4 & 0.05 & 3600 & 0.075 & $1.52^{+0.04}_{-0.03}$ & 23.1\\
		M2129 & 01833 & 2.29 & 21:29:27.054 & -07:42:35.72 & 06/12/17 & LGS & Kn3 & 0.05 & 2700 & 0.129 & $1.56^{+0.08}_{-0.05}$ & 22.1\\
		\hline
	\end{tabular}%
}
\vspace{0.5cm}
\label{tab:data_table}
\end{table*}

We are motivated to probe intermediate redshifts (z $\sim$ 1-3) from both an observational and theoretical standpoint. Observationally, this epoch corresponds to a peak in star formation rate (SFR) density (see \citealt{Madau_Dickinson_2014}), and thus galaxy assembly activity, where stellar feedback (dominated by young, massive stars) is most active. Similarly, models predict that the strength of galaxy coring can correlate with gas accretion rates and SFRs, so higher redshifts are likely more interesting to test this scenario.

\begin{figure}
  \includegraphics[width=\linewidth]{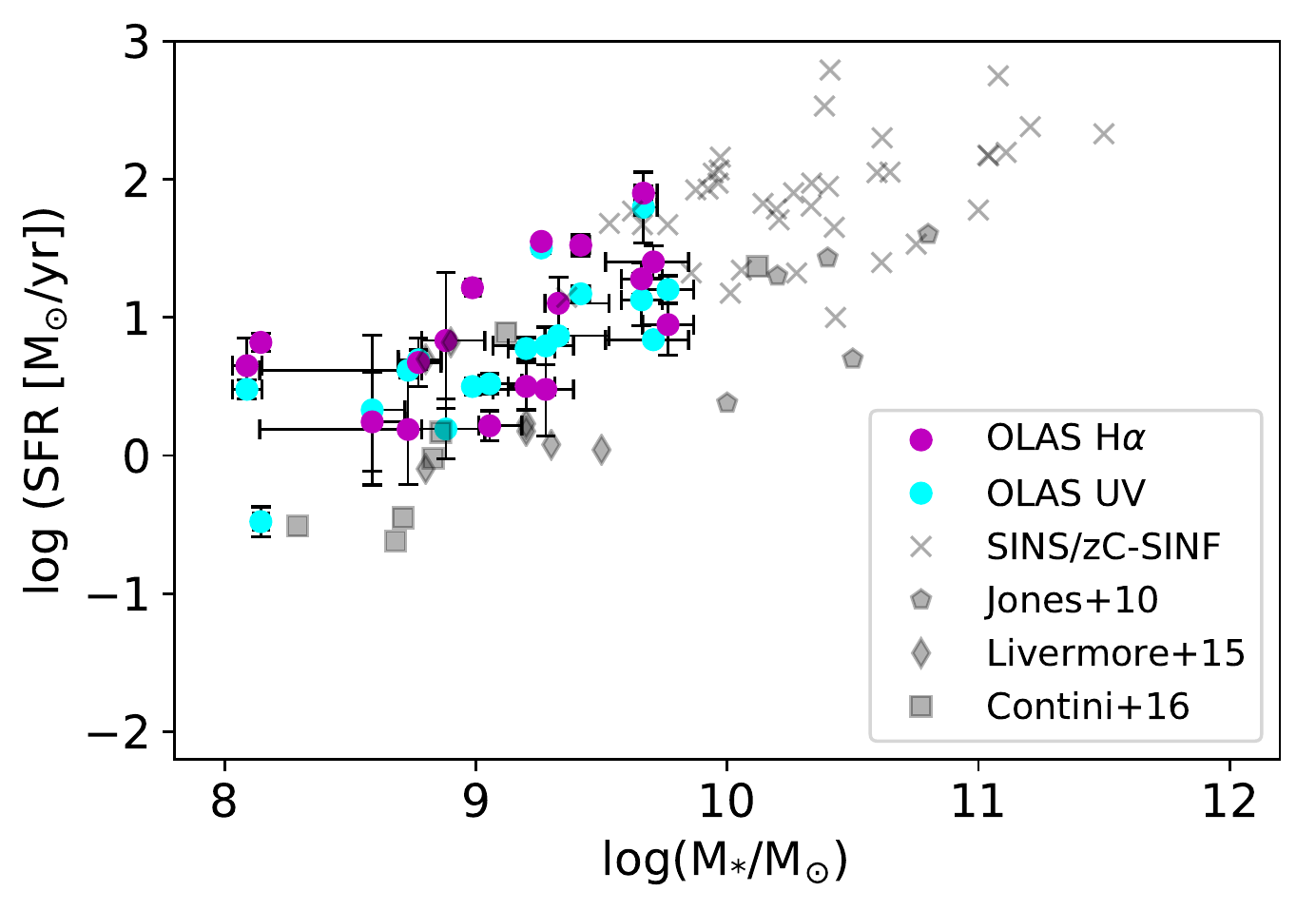}
  \caption{OLAS stellar masses and SFRs (shaded circles with error bars), with SFR measured from nebular emission lines (purple) and UV-continuum (cyan) for each source, reaching the mass and redshift range where feedback is predicted to have the strongest dynamical effect (M$_* \lesssim 10^9 $M$_{\odot}$). Gray data points show similar surveys with varying combinations of AO and lensing (crosses - AO only; \citealt{FS_2018} (SINS/zC-SINF), pentagons - both AO and lensing; \citealt{Jones_2010}, diamonds - both AO and lensing; \citealt{Livermore_2015}, squares - neither AO nor lensing, \citealt{Contini_2016}, z $>$ 1).}
  \label{fig:mass_vs_SFR}
\end{figure}

To gain a comprehensive understanding of the impact of stellar feedback on our universe at all epochs, we must discern between these different feedback models. To effectively study density profiles on sub-kpc scales needed to characterize dwarf galaxy structure, we require observations utilizing both adaptive optics (AO) and gravitational lensing. In this paper, we introduce a new survey, the OSIRIS Lens-Amplified Survey (OLAS), which takes advantage of massive galaxy clusters as gravitational lenses as well as AO at Keck Observatory with the OH-Suppressing Infrared Imaging Spectrograph (OSIRIS), pushing down to roughly 1.5 orders of magnitude lower in both stellar mass and SFR than other similar AO studies at intermediate redshifts (\citealt{FS_2018}), and dramatically extending the sample at the lowest masses for which lensing is necessary (\citealt{Stark_2008}, \citealt{Jones_2010}, \citealt{Livermore_2015}, \citealt{Contini_2016}), as seen in Figure \ref{fig:mass_vs_SFR}. Additionally, the AO corrections yield an order of magnitude increase in spatial resolution compared to seeing-limited kinematic studies (e.g. \citealt{Wuyts_2016}, \citealt{Stott_2016}). Typical seeing-limited resolution of $\gtrsim$ 0.5'' is insufficient to resolve even our lensed galaxy targets, whereas we achieve good spatial sampling from the combination of AO and lensing magnification. Future papers will further explore the resolved kinematics and dynamical mass profiles. Here, we focus on an initial test which is straightforward and theoretically well-motivated.

In this paper we compare galaxy kinematics with sSFRs to test observational signatures of core formation found in simulations. Physically, bursts of star formation drive powerful outflows which lower the central mass density in galaxies. This rapid change in gravitational potential is expected to drive both baryons and dark matter to wider orbits, manifesting as a lower stellar velocity dispersion, $\sigma_*$, with a correlation between $\sigma_*$ and recent sSFR (\citealt{ElBadry_2017}). The velocity dispersion of stars is a direct indicator of potential fluctuations; here we study the integrated velocity dispersion of H$\alpha$ in star forming regions (denoted in this paper as $\sigma$, unless otherwise stated). Unlike stars, gas kinematics can be altered due to stellar feedback, regardless of whether or not the gravitational potential is significantly affected. Nonetheless, we are testing a predicted correlation between $\sigma$ and sSFR, which is a necessary, but not sufficient, prerequisite for coring.

\begin{table*}
\centering
\caption{Data Properties}
\resizebox{0.9\textwidth}{!}{%
	\begin{tabular}{c c c c c c c c c c}
		\hline
		Cluster & ID &  Mass & log(SFR) (UV) & log(SFR) (H$\alpha$) & Local $\sigma$ & Integrated $\sigma$ & $\Delta v$\footnote{$\Delta$v is defined here as $v_{max} - v_{min}$} & $v/\sigma$ & Kinematic Class\footnote{1: smooth, monotonic velocity gradient, 2: disturbed kinematics.}\\
		 \ & \ & log($M_*/M_{\odot}$) & log($M_{\odot}/$yr) & log($M_{\odot}$/yr) & km/s & km/s & km/s & & \\
		\hline
		A370 & 02056 & $8.77^{+0.09}_{-0.08}$ & $0.69^{+0.06}_{-0.07}$ & $0.67^{+0.08}_{-0.18}$ & 11.7 $\pm$ 2.8 & 33.8 $\pm$ 4.5 & 36 & 1.53 $\pm$ 1.25 & 2\\
		A370 & 03097 & $9.32^{+0.20}_{-0.05}$ & $0.87^{+0.04}_{-0.04}$ & $1.10^{+0.20}_{-0.19}$ & 31.8 $\pm$ 2.5 & 68.8 $\pm$ 8.5 & 134 & 2.09 $\pm$ 0.275 & 1\\
		M0717 & 01828 & $8.72^{+0.05}_{-0.59}$ & $0.61^{+0.05}_{-0.05}$ & $0.19^{+0.21}_{-0.40}$ & 44.2 $\pm$ 7.2 & 50.7 $\pm$10.5 & 54 & 0.61 $\pm$ 0.224 & 2\\
		M0717 & 02064 & $8.08^{+0.06}_{-0.06}$ & $0.48^{+0.07}_{-0.07}$ & $0.65^{+0.13}_{-0.20}$ & 42.9 $\pm$ 5.3 & 74.4 $\pm$ 8.2 & 142 & 1.64 $\pm$ 0.288 & 1\\
		M0744 & 00920 & $9.05^{+0.13}_{-0.04}$ & $0.52^{+0.07}_{-0.07}$ & $0.21^{+0.11}_{-0.11}$ & 56.4 $\pm$ 1.4 & 65.5 $\pm$ 2.7 & 262 & 2.32 $\pm$ 0.189 & 1\\
		M0744 & 01203 & $9.26^{+0.01}_{-0.01}$ & $1.50^{+0.03}_{-0.04}$ & $1.55^{+0.03}_{-0.03}$ & 73.4 $\pm$ 3.1 & 97.9 $\pm$ 6.5 & 188 & 1.28 $\pm$ 0.128 & 1\\
		M0744 & 02341 & $9.19^{+0.11}_{-0.07}$ & $0.78^{+0.07}_{-0.08}$ & $0.50^{+0.16}_{-0.17}$ & 62.7 $\pm$ 0.99 & 47.9 $\pm$ 10.9 & 32 & 0.248 $\pm$ 0.150 & 2\\
		M1149 & 00593 & $9.41^{+0.03}_{-0.03}$ & $1.17^{+0.02}_{-0.02}$ & $1.52^{+0.07}_{-0.08}$ & 36.3 $\pm$ 2.7 & 82.6 $\pm$ 8.0 & 158 & 2.17 $\pm$ 0.373 & 1\\
		M1149 & 00683 & $8.14^{+0.03}_{-0.03}$ & $-0.48^{+0.09}_{-0.11}$ & $0.82^{+0.06}_{-0.07}$ & 43.2 $\pm$ 3.5 & 62.9 $\pm$ 9.9 & 102 & 1.17 $\pm$ 0.315 & 1\\
		M1149 & 01058 & $8.98^{+0.03}_{-0.03}$ & $0.50^{+0.01}_{-0.01}$ & $1.22^{+0.06}_{-0.04}$ & 23.4 $\pm$ 2.2 & 51.3 $\pm$ 5.1 & 46 & 0.965 $\pm$ 0.337 & 2\\
		M1149 & 01802 & $9.70^{+0.14}_{-0.19}$ & $0.84^{+0.02}_{-0.02}$ & $1.40^{+0.09}_{-0.12}$ & 55.4 $\pm$ 2.8 & 84.2 $\pm$ 4.2 & 264 & 2.39 $\pm$ 0.255 & 1\\
		M1423 & 00248 & $9.76^{+0.10}_{-0.10}$ & $1.20^{+0.08}_{-0.10}$ & $0.95^{+0.16}_{-0.22}$ & 26.5 $\pm$ 5.6 & 48.8 $\pm$ 11.9 & 52 & 0.968 $\pm$ 0.575 & 2\\
		M2129 & 00465 & $9.65^{+0.08}_{-0.08}$ & $1.13^{+0.13}_{-0.19}$ & $1.28^{+0.24}_{-0.12}$ & 63.4 $\pm$ 2.2 & 80.2 $\pm$ 6.0 & 84 & 0.655 $\pm$ 0.206 & 2\\
		M2129 & 00478 & $8.58^{+0.13}_{-0.01}$ & $0.33^{+0.23}_{-0.54}$ & $0.24^{+019}_{-0.36}$ & 22.0 $\pm$ 4.2 & 37.4 $\pm$ 10.1 & 20 & 0.469 $\pm$ 0.359 & 2\\
		M2129 & 01408 & $8.88^{+0.16}_{-0.09}$ & $0.19^{+0.15}_{-0.22}$ & $0.83^{+0.09}_{-0.49}$ & 64.4 $\pm$ 1.8 & 53.8 $\pm$ 8.1 & 54 & 0.425 $\pm$ 0.150 & 2\\
		M2129 & 01665 & $9.27^{+0.11}_{-0.21}$ & $0.79^{+0.10}_{-0.14}$ & $0.48^{+0.17}_{-0.34}$ & 59.0 $\pm$ 0.6 & 40.5 $\pm$ 17.4 & 4 & 0.03 $\pm$ 0.185 & 2\\
		M2129 & 01833 & $9.66^{+0.05}_{-0.03}$ & $1.80^{+0.16}_{-0.26}$ & $1.90^{+0.06}_{-0.05}$ & 79.1 $\pm$ 4.5 & 104.5 $\pm$ 13.0 & 158 & 1.01 $\pm$ 0.165 & 2\\
		\hline
	\end{tabular}%
	\label{tab:data_prop}
	}
\end{table*}

The paper is organized as follows. In Section \ref{sec:dataacq} we introduce our target selection criteria and sample. Section \ref{sec:methods1} describes our reduction and data extraction methods, both observationally and from the simulations. In Section \ref{sec:analysis}, we discuss the kinematic properties of our sample, present a relationship between $\sigma$, M$_*$, and sSFR from simulations and compare our data with the predicted relationship from the FIRE simulations. In Section \ref{sec:discussion}, we discuss our results. Finally, in Section \ref{sec:conclusions}, we summarize our results and discuss the possibilities this work proposes for future research. We assume a flat $\Lambda$CDM cosmology with H$_0$ = 70 km s$^{-1}$Mpc$^{-1}$, $\Omega_m$ = 0.3, and $\Omega_{\Lambda}$ = 0.7. 

\section{Data Acquisition} \label{sec:dataacq}

In order to probe stellar feedback at sub-kpc scales in both this critical range of low mass (log$(M_*/M_{\odot}$) $<$ 9) and high redshift (z $>$ 1), we require high-resolution spatial sampling and kinematic data. For our observations, we use OSIRIS (\citealt{Larkin_2006}), a near-infrared integral field unit (IFU) spectrograph at Keck Observatory with AO, to give us the necessary resolution and spatial sampling for detailed kinematic measurements at the scales of galactic cores.

Our targets are drawn from the Hubble Space Telescope (HST) program, the Grism Lens-Amplified Survey from Space (GLASS; \citealt{Treu_2015}). GLASS obtained near-IR grism spectroscopy of 10 galaxy clusters, detecting spatially resolved emission lines from large samples of background lensed galaxies. GLASS data and other HST imaging provide stellar masses, emission line maps, SFRs and metallicities (as presented in \citealt{Wang_2017}, \citealt{Wang_2018}). OSIRIS followup is critical as HST lacks the spectral resolution to measure kinematics. The GLASS data are necessary to pre-select targets with suitable stellar masses and redshifts. We distinctly look for galaxies with sufficiently strong H$\alpha$ for spatial mapping with OSIRIS that probe the aforementioned critical stellar mass and redshift range to test against simulations. However, this introduces a possible selection effect because our observations are biased towards galaxies with higher sSFRs (stronger H$\alpha$) whereas the simulations we compare to are not selected based on nebular emission strength. As we note in Section \ref{sec:sSFR}, our observations indeed have higher sSFRs than the simulated galaxies. A larger sample of observational data which pushes to lower sSFR will make this comparison more robust.

Between October 2016 and October 2018, we observed 17 gravitationally lensed, star forming galaxies with redshifts $1.2 < z < 2.3$ and  masses $8 < $ log$(M_*/M_{\odot}) < 9.8$, one of which was found to be an AGN and is excluded from this analysis. We observed in the H and K bands where the Strehl ratio is good, at both the 0.05 and 0.1'' pixel scale. The targets are mostly observed with laser guide star (LGS) AO, with some targets observed in the natural guide star (NGS) AO mode during laser failures. The OLAS sample to-date has a median FWHM of 0.136'' (Table \ref{tab:data_prop}), observed with clear sky conditions. A more detailed description of our target sample and relevant properties can be found in Tables \ref{tab:data_table} and \ref{tab:data_prop}. 

\section{Methods}
\label{sec:methods1}

\subsection{OSIRIS spectroscopic data}
\label{sec:methods}

\begin{figure*}
\centering
  \includegraphics[width=\linewidth]{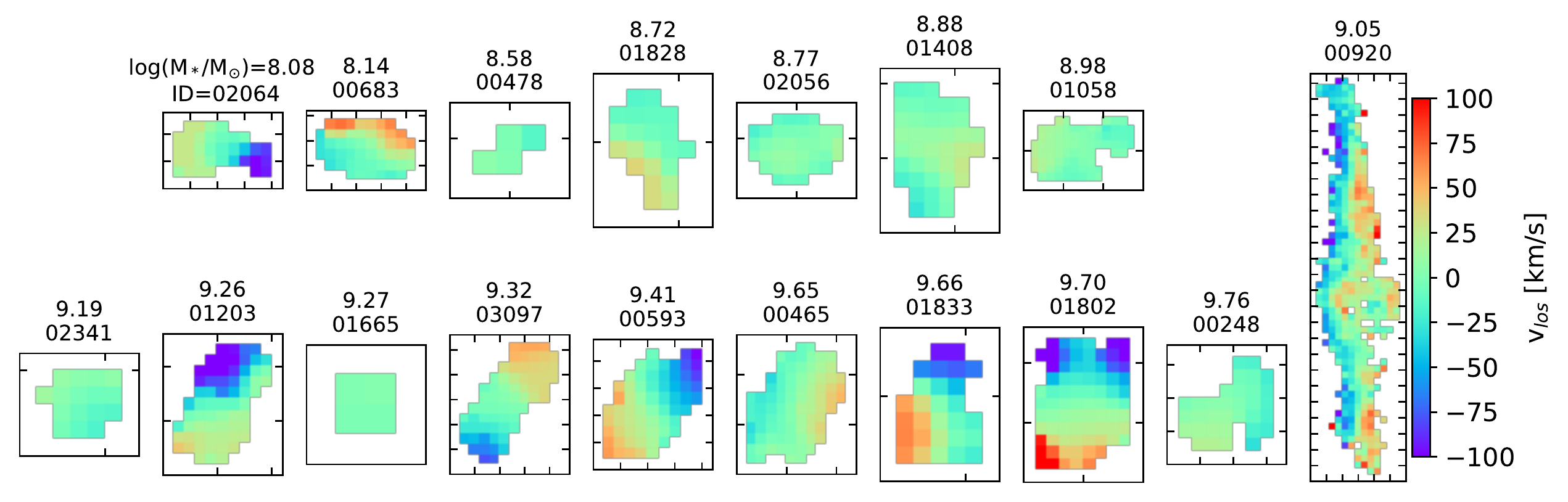}
  \caption{H$\alpha$ velocity maps for 17 star forming galaxies in our sample observed to date. The galaxies are sorted by stellar mass and are shown with a common velocity scale. Axis tick marks are 0.25 arcseconds in all panels. Targets with $\log(M_*/M_{\odot}) < $9 (top row) typically have lower velocity shear, whereas the more massive galaxies on the bottom row predominantly show ordered rotational motion.}
  \label{fig:kin_maps}
\end{figure*}

OSIRIS provides spatially resolved kinematic information in the form of a 3-D data cube. We reduced the data using the OSIRIS reduction pipeline\footnote{https://github.com/Keck-DataReductionPipelines/OsirisDRP} with a scaled sky subtraction. We perform a sigma-clipping to remove outlier pixels and smooth the data in the spatial directions to increase the signal to noise, yielding a larger area for our kinematic analysis. We then fit a gaussian in velocity space to the H$\alpha$ emission line at each spatial pixel (spaxel) to determine the velocity and local velocity dispersion, imposing a 5$\sigma$ detection threshold for acceptable fits (Figure \ref{fig:kin_maps}, Appendix \ref{app:kin_maps}).

We correct the local velocity dispersion (the mean value of $\sigma$ from individual spaxels) for beam smearing, determined from local velocity shear and the spatial PSF (smoothed according to each science target), as well as instrumental broadening, measured from skylines. In each spaxel, the observed velocity dispersion is corrected by subtracting contributions from both effects in quadrature, yielding a combined median correction of 22.2 km/s (median percent difference of 38.2\%). This method corrects for beam smearing due to all resolved velocity shear, regardless of whether the source exhibits rotation or disordered kinematics. The intrinsic local dispersion may still be overestimated if there is unresolved velocity structure, but we adopt this as a conservative approach with no dependence on modeling assumptions.

We also sum all spaxels into a 1-dimensional spectrum to measure the integrated velocity dispersion for each target. We correct the integrated velocity dispersion measurements for instrumental broadening, subtracting the instrumental contribution from the integrated dispersion in quadrature, yielding a median correction of 10.2 km/s (median percent difference of 15.5\%). This allows us to directly compare measured data with the simulations via the framework of \cite{ElBadry_2017}. The values for local and integrated velocity dispersions are listed in Table \ref{tab:data_prop}.

\subsection{Photometry and SED fitting}
We determine stellar masses of our targets by modeling their spectral energy distributions (SEDs), following similar methods as our previous work with the GLASS survey (e.g., \citealt{Jones_2015}; \citealt{Wang_2018}). HST photometry from the HFF (Hubble Frontier Fields; \citealt{Lotz_2017}) and CLASH (Cluster Lensing and Supernova Survey with Hubble; \citealt{Postman_2012}) surveys is fit with the stellar population synthesis code FAST (\citealt{Kriek_2009}). We adopt \cite{Bruzual_2003} spectral templates, Chabrier IMF, solar metallicity, \cite{Calzetti_2000} dust attenuation curve, and an exponentially declining star formation history. Each target's SED is well sampled with at least 7 bands of HST photometry, spanning observed wavelengths 0.4-1.6 $\mu$m. In each case we subtract the contribution of strong emission lines [OII], [OIII], H$\beta$, H$\alpha$+[NII], and [SII] from the broad-band continuum (following \citealt{Jones_2015}; emission lines are measured from GLASS data). This correction reduces the derived stellar masses by $\sim0.1$ - 0.2 dex, with typically larger effects in lower mass galaxies. Resulting best-fit stellar masses are given in Table \ref{tab:data_prop}.

For this work we are interested in determining SFRs averaged over the last $\sim$100 Myr, which we measure from the rest-frame UV continuum. This provides a value which is independent of the H$\alpha$-based SFR and probes a longer timescale. We measure UV spectral slopes $\beta$ (where $f_{\lambda} \propto \lambda^{\beta}$) and luminosities using HST F475W and F606W photometry. These filters correspond to rest-frame $\lambda = 1400-2600$ \AA\ for our targets. We use $\beta$ to correct for UV extinction following \cite{Meurer_1999}, and calculate SFRs from the UV luminosity following \cite{Kennicutt_1998} adjusted for a Chabrier IMF. The results are given in Table \ref{tab:data_prop} as SFR (UV). SED modeling also provides estimates of the recent SFR, which are generally consistent. We adopt the direct measurement of SFR from UV continuum as it is insensitive to the assumed star formation history used to model the SEDs.

The stellar masses and SFRs (listed in Table \ref{tab:data_prop}) in this paper are all corrected for lensing magnification ($\mu$; Table \ref{tab:data_table}), with magnification errors included in the uncertainties of the mass estimates. The lens models for the galaxy clusters used in this work were constructed using the lens modeling code Strong and Weak Lensing United (\citealt{Bradac_2005}; \citealt{Bradac_2009}). The code uses strongly and weakly lensed galaxies to reconstruct the gravitational potential on a refined pixel grid. This type of code is often referred to as free-form in that it does not fit the parameters of an analytical mass profile. Lens models for some of the clusters are described in previous works: A370 (\citealt{Strait_2018}), MACS1149 (\citealt{Finney_2018}), MACS1423 (\citealt{Hoag_2017}), MACS2129 (\citealt{Huang_2016}). The lens models for the remaining clusters are described in Hoag et al. (2018, in preparation).

\subsection{Comparison to simulations}
\label{sec:sims}

The main objective of this paper is to test whether trends between sSFR and H$\alpha$ gas velocity dispersion, $\sigma$, predicted by cosmological simulations are in fact present in our data. We compare our results with the Feedback in Realistic Environments (FIRE) simulations, for which galaxies at these masses form dark matter cores via stellar feedback, as a step towards understanding how stellar feedback drives gas kinematics and thus may drive galaxy-wide potential fluctuations and dark matter coring. An overview of the simulated galaxies used in this paper are listed in Table \ref{tab:sims}, and full details of the simulations are described in \cite{Hopkins_2018} and \cite{ElBadry_2018}.

\cite{ElBadry_2017} measured correlations between stellar kinematics and sSFRs. Given that we most readily measure gas kinematics via nebular emission lines, we recalculate for the FIRE simulations, carefully matching velocity dispersion calculations from FIRE to our observed data. Specifically, we measure H$\alpha$ emission in the observations, so we compare to the HII regions in the simulations, i.e. ionized gas near young stars. In every snapshot (40 snapshots per galaxy), we isolate the ionized gas particles within 100 pc of young ($<$10 Myr) stars and calculate the density weighted dispersion of their velocities, taking the median value over 100 randomized lines of sight. We do not account for the thermal contribution to $\sigma$ in our calculations, but they are negligible at these velocity dispersions.

Stellar masses and sSFRs are also determined for each simulated galaxy snapshot. As shown in Table \ref{tab:sims}, the simulated galaxies probe the same mass range as our observations. We calculate sSFR averaged over the last 10 (100) Myr by summing the total mass of stars younger than 10 (100) Myr within a given galaxy, corrected for stellar mass loss, and divide by that same timescale. 

\begin{figure}
  \includegraphics[width=\linewidth]{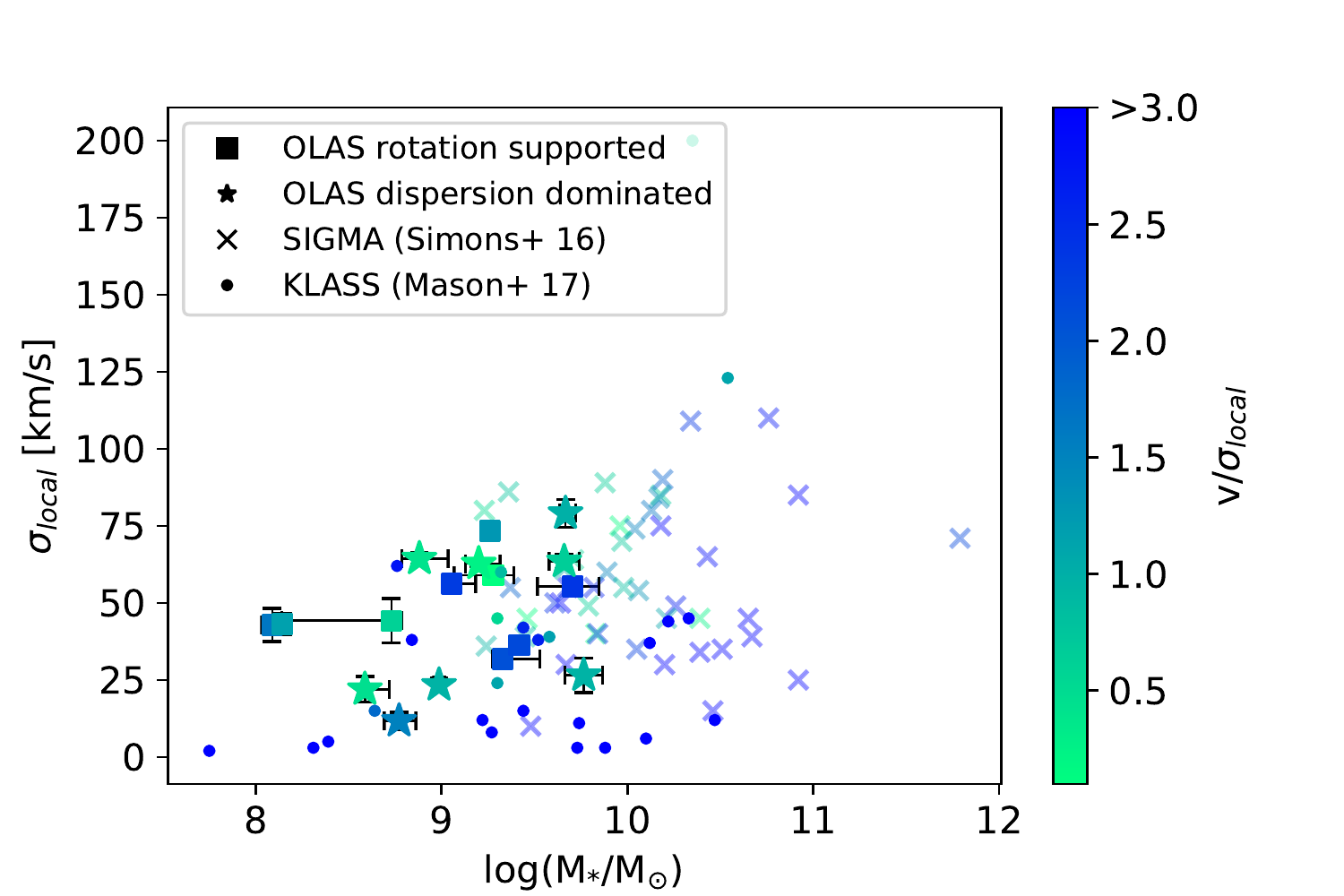}
  \caption{Local velocity dispersion plotted against stellar mass, color coded by $v/\sigma_{local}$ (Table \ref{tab:data_prop}). Filled squares represent OLAS targets with smooth velocity gradients while stars represent those that exhibit disordered kinematics. Crosses show data from the SIGMA survey (\citealt{Simons_2016}) and filled circles are data from the KLASS survey (\citealt{Mason_2017}). The $v/\sigma_{local}$ is computed differently within the KLASS sample, but we show their values for comparison. We do not see any trend in $v/\sigma_{local}$ with M$_*$ in the OLAS sample, giving no information of a mass scale where stellar feedback destroys disks.}
  \label{fig:v_over_sigma}
\end{figure}

\section{Analysis} \label{sec:analysis}

\subsection{Kinematic Maps}
\label{sec:kin_maps}

We present the lensed (image plane) velocity maps for our sample in Figure \ref{fig:kin_maps}. These maps are generated by plotting the velocity values at every spaxel, extracted as described in Section \ref{sec:methods}. Targets with $\log(M_*/M_{\odot}$) $<$ 9 typically have lower velocity shear, whereas the more massive galaxies predominantly show strong, ordered rotational motion. More detailed kinematic information (including velocity, velocity dispersion and error maps, HST images and H$\alpha$ maps) for each target are presented in Appendix \ref{app:kin_maps}.

Energetic bursts from stellar feedback episodes may also be strong enough to disrupt gas in galaxies; recent works show stellar feedback may flatten (or in some cases, invert) metallicity gradients and destroy gaseous disks (e.g., \citealt{Gibson_2013}; \citealt{Leethochawalit_2016}; \citealt{Ma_2017}; \citealt{Wang_2018}). This provides yet another test where OLAS is useful to compare to different feedback models. We separate our targets into two kinematic classes: (1) those that exhibit a smooth, monotonic velocity gradient, have $v/\sigma>1.1$, and available stellar mass map from HST imaging shows no clear signs of merging/disturbance, and (2) those that do not exhibit smooth gradients, have  $v/\sigma < 1.1$ or the available stellar mass map shows signs of merging/disturbance. Our choice of $v/\sigma$ threshold is physically motivated as follows: Rotation supports most of the dynamical mass when $v_{rot} > \sigma\sqrt{2}$ (as measured at the scale radius; \citealt{Burkert_2010}). In addition, the mean measured rotational velocity of randomly oriented disks will be reduced by a factor of $sin(i)=\pi/4$. Combined, these results yield a cutoff between rotationally supported and dispersion dominated systems with $v/\sigma=1.1$. This preliminary kinematic classification is to note if there are any immediate trends of galaxy morphology (i.e., rotating disks) with stellar mass, in search of a stellar mass range where stellar feedback may be strong enough to destroy disks.

Figure \ref{fig:v_over_sigma} shows the relationship between $\sigma_{local}$, $v$/$\sigma_{local}$, and M$_*$, where $v$ is the peak-to-peak velocity shear ($v$ = $\frac{1}{2} (v_{max} - v_{min}$)), and $\sigma_{local}$ is the mean local (measured per pixel) velocity dispersion of the galaxy (Table \ref{tab:data_prop}). To examine trends over a wider mass range, we additionally consider the results from the SIGMA survey (\citealt{Simons_2016}) and the KLASS survey (\citealt{Mason_2017};  we note that the velocity dispersions measured in KLASS being systematically lower than OLAS are likely due to a difference in modeling, particularly in the presence of unresolved velocity shear. Additionally, the samples are small and selection criteria differed between the surveys). If feedback destroys gaseous disks below some threshold mass, we would expect a sharp decrease in $v/\sigma_{local}$ and/or larger scatter below that stellar mass. We do not see a clear signature of this effect in Figure \ref{fig:v_over_sigma}. There is a slight positive correlation between $\sigma_{local}$ with M$_*$ ($r^2 = 0.28$) in the OLAS data. This correlation indicates that velocity dispersion is likely a source of dynamical support at low mass. However we find no evidence of a threshold mass for rotational disk support in the current analysis. This will be explored further in future work with more detailed kinematic classification and disk model fitting.

\subsection{Relationship between integrated velocity dispersion, mass, sSFR}
\label{sec:sSFR}

\cite{ElBadry_2017} discuss the relationship between sSFR and stellar velocity dispersion at fixed M$_*$. We expect qualitatively similar trends for our calculations of velocity dispersion using ionized gas near young stars. To mirror these results, we separate our targets into near constant mass bins of log($M_*$/$M_{\odot}$) $\sim$ 8, 8.6, 8.8, 9.2, and 9.7, matching our observed galaxy masses to galaxies from the FIRE simulations, where all galaxies are within 0.2 dex of their respective mass bin. We compare the integrated velocity dispersion ($\sigma$) from OLAS to match our calculations from the FIRE simulations. By examining the relationship between sSFR and $\sigma$ at fixed stellar mass, we establish a physical link between kinematics and sSFR (i.e. due to feedback-driven outflows), isolated from mass-dependent trends.

Compiling all of our simulated galaxies ranging in stellar masses $7.5 < $ log$(M_*/M_{\odot}) < 9.9$, redshifts $1.2< $ z $< 2.7$ and sSFRs $-11.75 < \log(sSFR_{neb}(M_{\odot}/yr)) < -7.9$, $-12.6 < \log(sSFR_{UV}(M_{\odot}/yr))  < -8.3$, we parametrize $\sigma_{pred}$ from the FIRE simulations (independent of the observed data) into the following general form, where $\sigma_{pred}$ = $\sigma_{pred}$(M$_*$, sSFR):

\begin{multline}
\log\sigma_{pred}= a\log(sSFR)+b\log(M_*)\\+c\log(M_*)\log(sSFR),
\label{eq:sim_fit}
\end{multline}
where M$_*$ and sSFR are in units of solar masses (M$_{\odot}$) and yr$^{-1}$. We find no dependence on redshift within the simulated galaxies; including redshift as a parameter does not significantly improve the fit. This is in agreement with the findings from \cite{Hung_2018} where they find intrinsic velocity dispersion within the FIRE simulations to be independent of redshift for z $>$ 1. We find the coefficients $a, b,$ and $c$ for both sSFR averaged over 10 and 100 Myr, which we compare to the H$\alpha$-based and UV-based SFR measurements, respectively, listed in Table \ref{tab:sim_fit}.

\begin{table}[h]
\centering
\caption{Fit Parameters From FIRE}
\begin{tabular}{c c c c c}
\hline
sSFR [Myr]\footnote{Timescale (in Myr) over which SFR is averaged} & a & b & c & 1$\sigma$ scatter\footnote{Rms (68\%) scatter in $\log \sigma_{pred}$ around the best fit line}\\
 \hline
10 & 0.1006 & 0.3892 & 0.0126 & 0.171\\
100 & 0.0732 & 0.4120 & 0.0181& 0.175
\end{tabular}%
\label{tab:sim_fit}
\end{table}

\begin{figure}[h]
  \includegraphics[width=\linewidth]{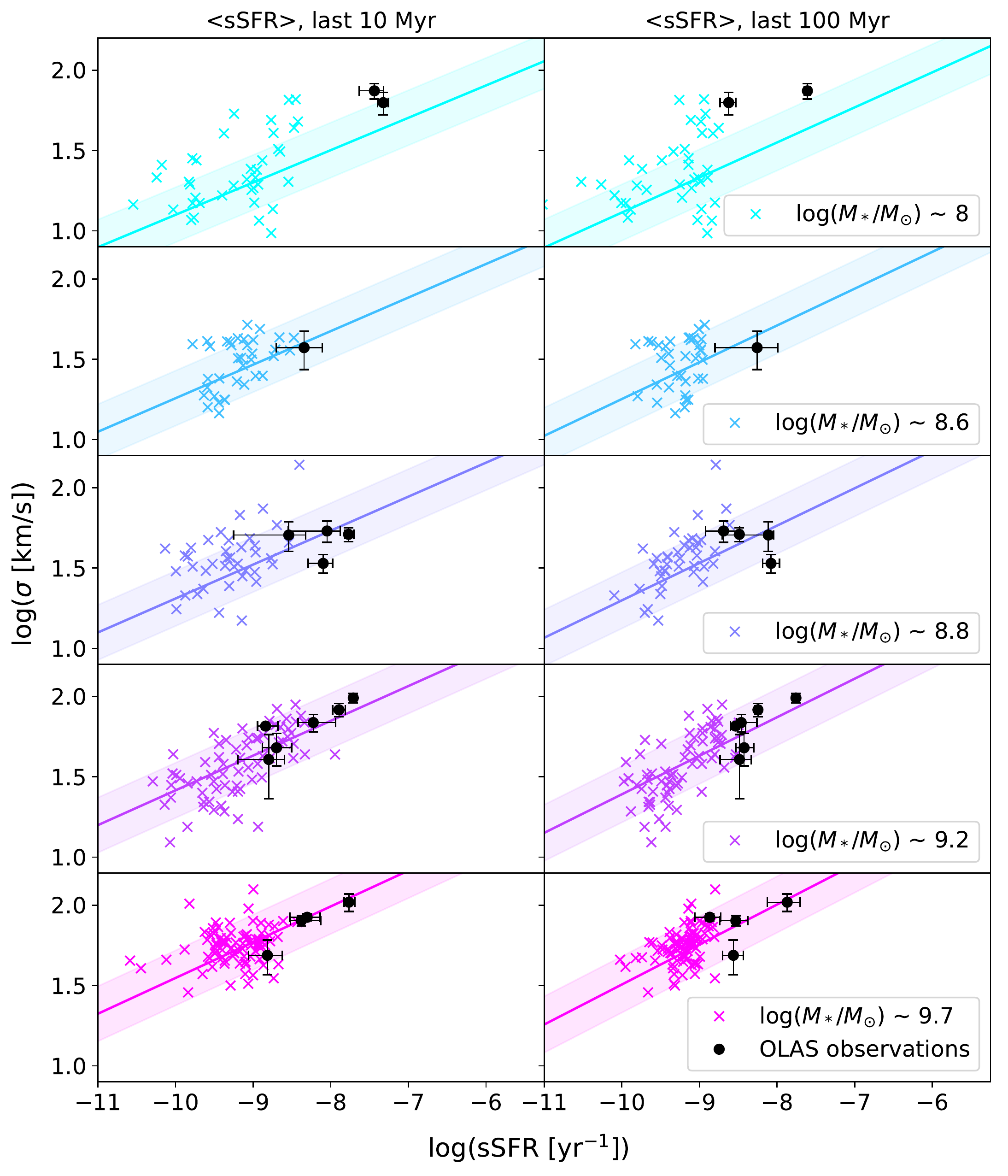}
  \caption{Colored panels show, at fixed mass, the relationships between integrated $\sigma$ and sSFR for sSFR averaged over both 10 (left) and 100 (right) Myr, as compared to observed line of sight velocity dispersion from our OSIRIS data (black points with error bars). Crosses represent extracted values from individual simulated galaxies. Solid lines and shaded regions show the dispersion and 1$\sigma$ scatter from Equation \ref{eq:sim_fit} evaluated in each panels respective galaxy mass using fitted parameters from Table \ref{tab:sim_fit}.}

  \label{fig:sSFR_vs_dispersion}
\end{figure}

The scenario for baryonic coring requires feedback to drive coherent gas outflows on the scale of the entire galaxy, thus causing an increase in $\sigma$ with increased sSFR. Figure \ref{fig:sSFR_vs_dispersion} shows the best-fit lines compared with both simulations and observed galaxies. At fixed stellar mass, we observe a clear trend in $\sigma$ with sSFR. This correlation quantitatively supports the hypothesis that star formation and associated feedback indeed have an effect on the gas kinematics, a prerequisite for feedback-driven core formation. However we note that such a correlation may arise from multiple processes; for example a high gas fraction could induce both higher sSFR and velocity dispersion (e.g., \citealt{Genzel_2011}). Nonetheless, equation \ref{eq:sim_fit} provides a good fit to the simulated galaxies over the full range of stellar mass and sSFR probed. The OLAS observations typically fall within the scatter of the fit. The observed galaxies have generally higher sSFR than the simulated sample, which could potentially introduce a small bias in extrapolating the fit. However we expect no significant effect on the conclusions, and if anything the observational data provide further support that the trend seen in simulations holds over a wide dynamic range. This could be further addressed in future work with an increased sample of simulated galaxies spanning a broader range of sSFR.

\subsection{Comparing data and simulations}

\begin{figure*}
\centering
  \includegraphics[width=0.9\linewidth]{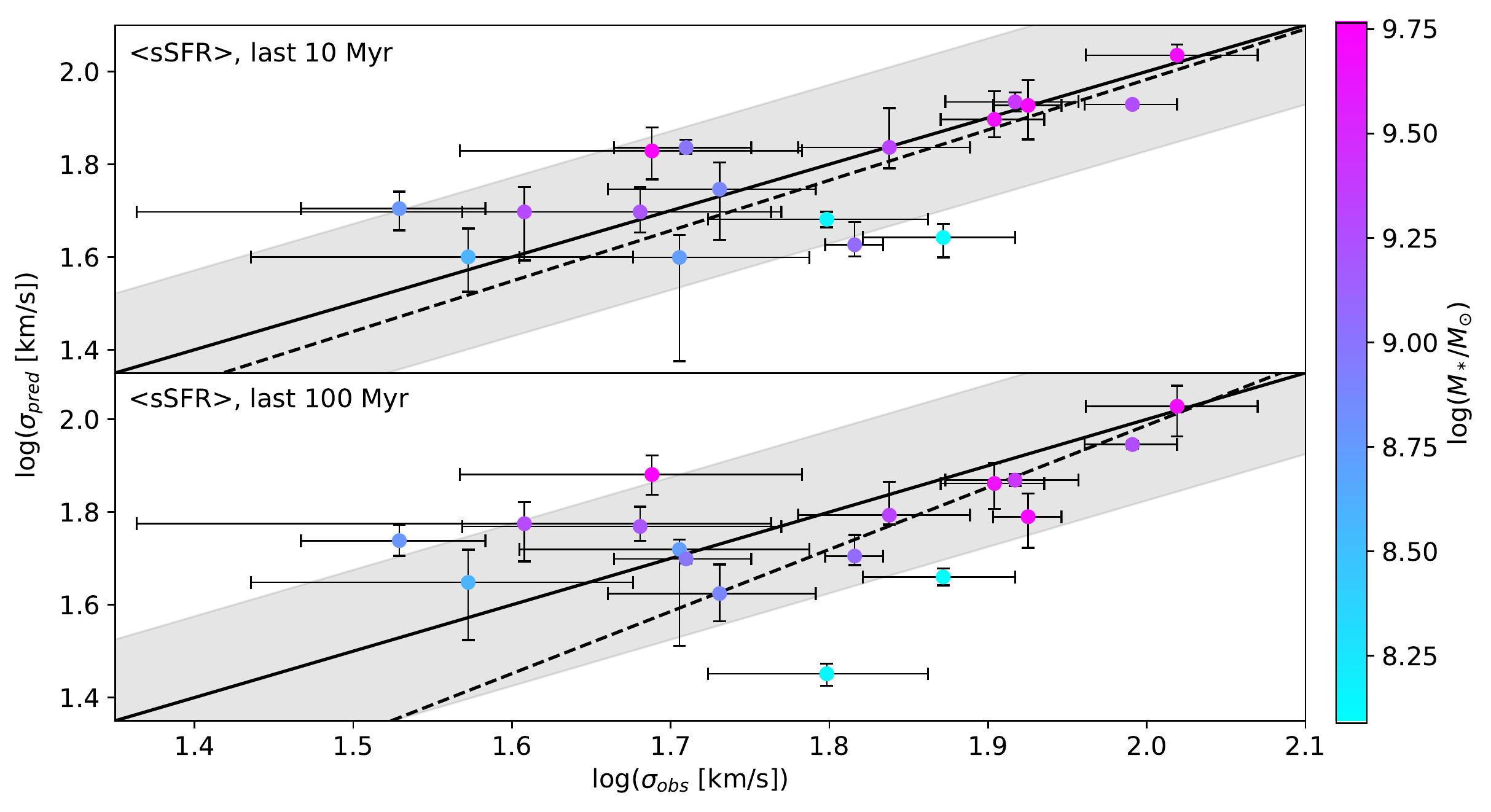}
  \caption{Measured integrated $\sigma$ ($\sigma_{obs}$) for each galaxy in our OSIRIS sample, compared to predictions from the simulations, $\sigma_{pred}$ (based on $M_*$ and sSFR, Equation \ref{eq:sim_fit}), color coded by galaxy stellar mass for sSFR averaged over 10 (top) and 100 (bottom) Myr. In both panels, the solid line represents $\sigma_{obs} = \sigma_{pred}$ with the shaded region showing 68\% scatter of simulated galaxies around the best fit to $\log\sigma_{pred}$. The dashed line shows the best fit line measured from the data points. Our observations of $\sigma$ agree best with the predicted velocity dispersion from the simulations using Equation \ref{eq:sim_fit}, when sSFR is calculated over a 10 Myr timescale rather than a 100 Myr timescale. Over 100 Myr timescales, gas kinematics are affected by processes that act over dynamical timescales, whereas over 10 Myr timescales gas kinematics are affected by recent episodes of star formation.}
  \label{fig:comparison}
\end{figure*}

We also compare our observed OLAS velocity dispersions to the predicted values from the simulations using Equation \ref{eq:sim_fit}, combining our data from Figure \ref{fig:sSFR_vs_dispersion} into one relationship (where $\sigma_{pred} = \sigma(M_*, sSFR)$, using SFR averaged over both 10 and 100 Myr). Figure \ref{fig:comparison} presents $\sigma_{pred}$ compared to OLAS observations. 16/17 observed galaxies are consistent within their 1$\sigma$ uncertainties to $\sigma_{pred}$ with both sSFR averaged over 10 and 100 Myr.

To quantify whether the observational data are consistent with the trend seen in simulations, we compute a linear fit of $\sigma_{obs}$ vs $\sigma_{pred}$. For both SFR averaged over both 10 and 100 Myrs, the results are consistent with expectations to within 1 standard deviation (i.e., unity slope and zero offset). The reduced chi-square values are 0.3 and 0.5 for sSFR averaged over 10 and 100 Myr, respectively, where values less than 1 may indicate observational uncertainties are overestimated, or that simulations over-predict the scatter by 30-50\%.

\section{Discussion} \label{sec:discussion}

\subsection{Rotational support at low mass}

We preliminarily explore the relationship between smooth velocity gradients and galaxy stellar mass in search of a mass scale where stellar feedback is strong enough to overcome galaxy gravitational potentials and destroy disks. At higher masses we expect feedback to have a less dynamical effect due to deep potential wells.

Recent works show that at z $\sim$ 2, high-mass galaxies ($\log(M_*/M_{\odot}) >$ 10.2) are generally rotationally supported with primitive disks, while large fractions of lower mass galaxies are just beginning to form disks (\citealt{Simons_2016}). We do not see any such trends within our current data set with galaxies across a wide stellar mass range showing velocity shears (as shown in Figure \ref{fig:v_over_sigma}). Within the FIRE simulations, there are not generally disk-like structures in low-redshift galaxies below $\log(M_*/M_{\odot}) \sim$ 10. As noted in \citealt{ElBadry_2018b}, the FIRE simulations may over-predict dispersion in low-mass galaxies at low redshift and thus skew predictions of rotational support to higher masses, although this trend has not been tested at the redshifts in this analysis. We require a larger sample, as well as a more complete kinematic analysis to give further insight on a mass scale at which feedback destroys disks in galaxies.

\subsection{sSFR and Dispersion}

If stellar feedback significantly alters the gravitational potentials of dwarf galaxies, we expect to observe a direct relationship between line-of-sight velocity dispersion and recent star formation (\citealt{ElBadry_2017}). We indeed find a direct relationship between $\sigma$ and sSFR: at fixed mass, higher sSFR correlates with higher $\sigma$, consistent with predicted values from simulated galaxies (Figure \ref{fig:comparison}; Equation \ref{eq:sim_fit}).

We examine this relationship over two relevant timescales, with sSFR averaged over 10 and 100 Myr. 10 Myr timescales trace the lifetime of massive stars, which are the main contributors to stellar feedback processes, whereas 100 Myr timescales cover dynamical timescales of galaxies, allowing the possibility for multiple starbursts episodes in addition to giving galaxies adequate time for their kinematics to respond to fluctuating gravitational potentials.

Our data exhibit a near equivalent correlation - 0.171 and 0.175  dex scatter - for sSFR averaged over both 10 and 100 Myr respectively, as shown in Table \ref{tab:sim_fit}. The simulated data also appear to have similar scatter in fixed mass bins, as shown in Figure \ref{fig:sSFR_vs_dispersion}. This differs from the analysis in \cite{ElBadry_2017}, where they show a stronger correlation between {\it stellar} kinematics and sSFR when sSFR is averaged over 100 Myr as compared to sSFR averaged over 10 Myr. This disagreement may be due to the differences in analyses; this work evaluates the velocity dispersion of {\it gas} in ionized HII regions near young stars, contrasting the direct dispersion of all stars as measured in \cite{ElBadry_2017}. Completing our analysis with velocity dispersion calculated from all stars in the simulations yields a tighter correlation (smaller scatter), with SFR averaged over 100 Myr, consistent with \cite{ElBadry_2017}. The stronger correlation in $\sigma$ of all stars with sSFR averaged over 100 Myr may be due to the dynamical time delay between rapidly fluctuating gravitational potentials and these fluctuations altering kinematics of stars. Gas kinematics may be affected on shorter dynamical timescales than stellar kinematics, as gas is affected by stellar feedback on short timescales whereas stellar kinematics are not.

Since our predicted $\sigma$s are affected by both sSFR and stellar mass (Equation \ref{eq:sim_fit}), we are careful to verify that the observed trend in dispersion is not solely dependent upon mass. The correlation coefficient of the OLAS sample between M$_*$ and integrated $\sigma$ is 0.34, smaller than the correlation between the observed and predicted $\sigma$: 0.69 and 0.47 for sSFR averaged over 10 and 100 Mry, respectively. Thus, $\sigma(M_*, sSFR)$ vs $\sigma$ is more strongly correlated than M$_*$  vs $\sigma$, showing that the intrinsic relationship between $M_*$ and $\sigma$ is not the only contributor to the observed correlation of $\sigma$ with sSFR. From this, we conclude that $\sigma$ and sSFR are indeed correlated, and this observed relation is in good agreement with the FIRE simulations.

\subsection{Implications for core formation}

Simulations predict that core formation in dwarf galaxies may be driven by stellar feedback, provided that two conditions are met. First, feedback-driven mass loss rates must be dynamically significant. Second, multiple such strong feedback events must occur over a galaxy's star formation history in order to eventually form a large central core. This study addresses the first condition, focusing on the effects of star formation feedback at a single epoch. To address core formation we must additionally consider dwarf galaxy star formation histories probed by complementary methods (e.g. \citealt{Teyssier_2013}, \citealt{ElBadry_2016}, \citealt{Sparre_2016}). These results notably show bursty star formation histories oscillating over timescales on the order of $\sim$ 100 Myr, with starbursts on the order of $\sim$ 10-100 Myr in cosmological simulations. As discussed in \cite{ElBadry_2017}, the power-law slopes of galaxy density profiles are predicted to be anti-correlated with sSFR and $\sigma$. In other words, cores and cusps oscillate in time with sSFR; periods of high sSFR and $\sigma$ yield cusp-y density profiles, which flatten into cores following starburst episodes with strong gaseous outflows.

Of course it is not possible to measure reliable SFR histories for individual galaxies on the timescales relevant to verify these oscillations. A promising observational test of burstiness is to compare the SFR measured on different timescales, e.g. from H$\alpha$ and UV continuum, in large samples. Recently \cite{Emami_2018} used this technique to study dwarf galaxies at z=0, finding galaxies with M$_* < 10^{7.5} M_{\odot}$ undergo strong bursts of star formation over shorter timescales ($t_{burst}<$ 30 Myr), while more massive galaxies ($M_*> 10^{8.5} M_{\odot}$) experience longer, less powerful bursts, with $t_{burst}>$ 300 Myr. These results support bursty star formation histories which appear to be required for core formation. A similar study at $z\simeq2$ would be of great interest in establishing the statistical properties of starburst timescales within the low-mass galaxy population. Differences between H$\alpha$ and UV-derived SFRs for the OLAS sample (Figure \ref{fig:mass_vs_SFR}) indeed suggest that SFR fluctuates on $<$ 100 Myr timescales, in our modest sample. However at present it remains an open question as to whether galaxies in fact undergo the number and frequency of repetitive star formation and feedback episodes needed to form cores, and over what timescales this transition may occur.

Our analysis shows that, using both fixed mass bins as well as a parameterized fit including stellar mass dependence, we observe the same trends in $\sigma$ with sSFR as in cosmological simulations. However this is only one piece of the puzzle. We are careful to note that our analysis traces gas kinematics in dwarf galaxies, and not stellar kinematics. Gas kinematics of HII regions are affected by both feedback-induced outflows and the galaxy gravitational potential, unlike the kinematics of stars which are solely dependent upon the potential. The agreement of gas dispersion with sSFR is a precursor of the core formation process; if we did not observe this trend in the gas, it would be difficult to argue the trend would follow in the stars (and dark matter). While we cannot directly conclude stellar feedback resolves the cusp-core problem from these results, we can support that our observations of gas kinematics are in excellent agreement with predictions from simulations that do resolve the cusp-core problem with feedback. This initial observational test therefore paves the way for more stringent future kinematic studies such as dynamical mass modeling, and direct measurement of stellar kinematics with deep spectroscopy and/or future larger telescopes

\section{Summary}  \label{sec:conclusions}

The OSIRIS Lens-Amplified Survey (OLAS) is designed to address the role of stellar feedback in low mass galaxies at intermediate redshift. In this paper we introduced OLAS and presented the first results from the current sample of 17 galaxies. By using a combination of gravitational lensing and Keck AO, we provide a unique, spatially resolved kinematic sample of galaxies with low M$_*$ and sSFR, pushing $\sim$1.5 orders of magnitude lower in both quantities than other AO surveys (Figure \ref{fig:mass_vs_SFR}), and explored tests of feedback driven core formation in low mass galaxies. Our main results are as follows:

\begin{enumerate}
\item We presented spatially resolved, image plane H$\alpha$ velocity maps (Figure \ref{fig:kin_maps}) and preliminarily explored the relationship between rotational support ($v/\sigma_{local}$), ordered vs disordered kinematics, and stellar mass in comparison with recent literature \citep[][Figure \ref{fig:v_over_sigma}]{Simons_2016}. We used this relationship as a proxy for the strength of stellar feedback; i.e., to examine whether stellar feedback is strong enough to disrupt the overall gas kinematics of galaxies and destroy disks at these stellar mass scales. In this rudimentary classification scheme, we do not note any convincing trends that would signify a mass threshold for disruption of gaseous disks. We require a larger data set, in combination with a comprehensive dynamical analysis for kinematic classification, in order to search for a stellar mass scale where feedback destroys disks in galaxies as expected from theoretical arguments.
\item We tested the relationship between velocity dispersion and sSFR as predicted in \cite{ElBadry_2017} from the FIRE simulations. While \cite{ElBadry_2017} analyzed this relationship using stellar velocity dispersions, we studied star forming HII regions of the simulations to best mimic our observations. From the FIRE simulations we quantified line-of-sight $\sigma$ as a function of M$_*$ and sSFR, presenting the functional form (Equation \ref{eq:sim_fit}) and its parameters (Table \ref{tab:sim_fit}), finding no significant redshift dependence. Our results are consistent with predictions from simulations to within one standard deviation using sSFRs averaged over both 10 and 100 Myr. We confirm that galaxies with higher sSFR have higher $\sigma$ at fixed mass, and the observed trend is in good agreement with that predicted by our analysis of theoretical simulations.
\item Simulations have predicted that core formation is driven by stellar feedback which alters the gravitational potentials, and thus the stellar kinematics and central dark matter density profiles. While we cannot make direct claims about the motions of stars with OLAS observations, this work uses the kinematics of ionized gas in star forming regions which are an appropriate tracer for stellar kinematics. We find that our observed gas kinematics are consistent with predictions from the FIRE simulations which do generate dark matter cores at these masses, suggesting that stellar feedback may indeed induce core formation in low mass galaxies as the simulations show. This agreement paves the way for further, more definitive, dynamical studies.
\end{enumerate}

This work presents a new observational test of the origin of the cusp-core problem, one of the most important and outstanding challenges to the cold dark matter paradigm. Answering whether or not cores form through baryonic feedback processes can either validate or disprove the CDM hypothesis. This new approach is interesting because by extending our observations out to higher redshift (instead of restricting to the local universe), we probe the epoch in cosmic history where stellar feedback has the most dynamical effect. If we could not observe the predicted trend between $\sigma$ and sSFR where feedback is predicted to have the strongest effect on galaxy kinematics, we would not expect feedback to be a catalyst for core formation.

We do find kinematic signatures of stellar feedback altering the kinematics of dwarf galaxies which are in good agreement with fully cosmological simulations that resolve the cusp-core problem with baryonic feedback. Future studies using JWST, TMT/IRIS and other thirty-meter class telescopes will allow us to observe galaxies with lower sSFRs, providing a wider dynamic range, as well as observe stellar kinematics, offering more direct tracers of galaxy gravitational potentials. These upcoming advances will further allow us to use stellar feedback in dwarf galaxies beyond the local group as a probe for studying dark matter core formation, helping us understand the major mass component of the universe.

\acknowledgments

We thank the anonymous referee for a constructive report which substantially improved the content and clarity of this manuscript. We thank Matthew Orr for productive discussions on comparing our observations to the FIRE simulations. Support for the Grism Lens Amplified Survey from Space (GLASS) (HST-GO-13459) was provided by NASA through a grant from the Space Telescope Science Institute (STScI). TJ acknowledges support from NASA through grants HST-HF2-51359 and HST-GO-15077 awarded by STScI. AW was supported by NASA through ATP grant 80NSSC18K1097 and grants HST-GO-14734 and HST-AR-15057 from STScI. The simulations were run using the Extreme Science and Engineering Discovery Environment (XSEDE) supported by NSF grant ACI-1548562, using Blue Waters via allocation PRAC NSF.1713353 supported by the NSF, and using the NASA HEC Program through the NAS Division at Ames Research Center. This research made use of Astropy,\footnote{http://www.astropy.org} a community-developed core Python package for Astronomy \citep{Astropy_2013, Astropy_2018}. The data presented herein were obtained at the W. M. Keck Observatory, which is operated as a scientific partnership among the California Institute of Technology, the University of California and the National Aeronautics and Space Administration. The Observatory was made possible by the generous financial support of the W. M. Keck Foundation. The authors wish to recognize and acknowledge the very significant cultural role and reverence that the summit of Maunakea has always had within the indigenous Hawaiian community.  We are most fortunate to have the opportunity to conduct observations from this mountain.

\bibliography{mybib}

\begin{thebibliography}{}
\expandafter\ifx\csname natexlab\endcsname\relax\def\natexlab#1{#1}\fi

\bibitem[{Akrami {et~al.}(2018)}]{Planck_2018}
Akrami, Y., {et~al.} 2018, arXiv:1807.06205

\bibitem[{{Astropy Collaboration} {et~al.}(2013){Astropy Collaboration},
  {Robitaille}, {Tollerud}, {Greenfield}, {Droettboom}, {Bray}, {Aldcroft},
  {Davis}, {Ginsburg}, {Price-Whelan}, {Kerzendorf}, {Conley}, {Crighton},
  {Barbary}, {Muna}, {Ferguson}, {Grollier}, {Parikh}, {Nair}, {Unther},
  {Deil}, {Woillez}, {Conseil}, {Kramer}, {Turner}, {Singer}, {Fox}, {Weaver},
  {Zabalza}, {Edwards}, {Azalee Bostroem}, {Burke}, {Casey}, {Crawford},
  {Dencheva}, {Ely}, {Jenness}, {Labrie}, {Lim}, {Pierfederici}, {Pontzen},
  {Ptak}, {Refsdal}, {Servillat}, \& {Streicher}}]{Astropy_2013}
{Astropy Collaboration}, {Robitaille}, T.~P., {Tollerud}, E.~J., {et~al.} 2013,
  \aap, 558, A33

\bibitem[{{Astropy Collaboration} {et~al.}(2018){Astropy Collaboration},
  {Price-Whelan}, {Sip{\H o}cz}, {G{\"u}nther}, {Lim}, {Crawford}, {Conseil},
  {Shupe}, {Craig}, {Dencheva}, {Ginsburg}, {VanderPlas}, {Bradley},
  {P{\'e}rez-Su{\'a}rez}, {de Val-Borro}, {Aldcroft}, {Cruz}, {Robitaille},
  {Tollerud}, {Ardelean}, {Babej}, {Bach}, {Bachetti}, {Bakanov}, {Bamford},
  {Barentsen}, {Barmby}, {Baumbach}, {Berry}, {Biscani}, {Boquien}, {Bostroem},
  {Bouma}, {Brammer}, {Bray}, {Breytenbach}, {Buddelmeijer}, {Burke},
  {Calderone}, {Cano Rodr{\'{\i}}guez}, {Cara}, {Cardoso}, {Cheedella},
  {Copin}, {Corrales}, {Crichton}, {D'Avella}, {Deil}, {Depagne}, {Dietrich},
  {Donath}, {Droettboom}, {Earl}, {Erben}, {Fabbro}, {Ferreira}, {Finethy},
  {Fox}, {Garrison}, {Gibbons}, {Goldstein}, {Gommers}, {Greco}, {Greenfield},
  {Groener}, {Grollier}, {Hagen}, {Hirst}, {Homeier}, {Horton}, {Hosseinzadeh},
  {Hu}, {Hunkeler}, {Ivezi{\'c}}, {Jain}, {Jenness}, {Kanarek}, {Kendrew},
  {Kern}, {Kerzendorf}, {Khvalko}, {King}, {Kirkby}, {Kulkarni}, {Kumar},
  {Lee}, {Lenz}, {Littlefair}, {Ma}, {Macleod}, {Mastropietro}, {McCully},
  {Montagnac}, {Morris}, {Mueller}, {Mumford}, {Muna}, {Murphy}, {Nelson},
  {Nguyen}, {Ninan}, {N{\"o}the}, {Ogaz}, {Oh}, {Parejko}, {Parley}, {Pascual},
  {Patil}, {Patil}, {Plunkett}, {Prochaska}, {Rastogi}, {Reddy Janga},
  {Sabater}, {Sakurikar}, {Seifert}, {Sherbert}, {Sherwood-Taylor}, {Shih},
  {Sick}, {Silbiger}, {Singanamalla}, {Singer}, {Sladen}, {Sooley},
  {Sornarajah}, {Streicher}, {Teuben}, {Thomas}, {Tremblay}, {Turner},
  {Terr{\'o}n}, {van Kerkwijk}, {de la Vega}, {Watkins}, {Weaver}, {Whitmore},
  {Woillez}, {Zabalza}, \& {Astropy Contributors}}]{Astropy_2018}
{Astropy Collaboration}, {Price-Whelan}, A.~M., {Sip{\H o}cz}, B.~M., {et~al.}
  2018, \aj, 156, 123

\bibitem[{{Bose} {et~al.}(2018){Bose}, {Frenk}, {Jenkins}, {Fattahi}, {Gomez},
  {Grand}, {Marinacci}, {Navarro}, {Oman}, {Pakmor}, {Schaye}, {Simpson}, \&
  {Springel}}]{Bose_2018}
{Bose}, S., {Frenk}, C.~S., {Jenkins}, A., {et~al.} 2018, ArXiv e-prints,
  arXiv:1810.03635

\bibitem[{Brada{\v c} {et~al.}(2009)Brada{\v c}, Treu, Applegate, Gonzalez,
  Clowe, Forman, Jones, Marshall, Schneider, \& Zaritsky}]{Bradac_2009}
Brada{\v c}, M., Treu, T., Applegate, D., {et~al.} 2009, The Astrophysical
  Journal, 706, 1201

\bibitem[{{Brada{\v c}} {et~al.}(2005){Brada{\v c}}, {Schneider, P.},
  {Lombardi, M.}, \& {Erben, T.}}]{Bradac_2005}
{Brada{\v c}}, M., {Schneider, P.}, {Lombardi, M.}, \& {Erben, T.} 2005, A\&A,
  437, 39

\bibitem[{Bruzual \& Charlot(2003)}]{Bruzual_2003}
Bruzual, G., \& Charlot, S. 2003, Monthly Notices of the Royal Astronomical
  Society, 344, 1000

\bibitem[{Buckley \& Fox(2010)}]{Buckley_Fox_2010}
Buckley, M.~R., \& Fox, P.~J. 2010, Phys. Rev. D, 81, 083522

\bibitem[{Calzetti {et~al.}(2000)Calzetti, Armus, Bohlin, Kinney, Koornneef, \&
  Storchi-Bergmann}]{Calzetti_2000}
Calzetti, D., Armus, L., Bohlin, R.~C., {et~al.} 2000, The Astrophysical
  Journal, 533, 682

\bibitem[{Chan {et~al.}(2015)Chan, Kereš, Oñorbe, Hopkins, Muratov,
  Faucher-Giguère, \& Quataert}]{Chan_2015}
Chan, T.~K., Kereš, D., Oñorbe, J., {et~al.} 2015, Monthly Notices of the
  Royal Astronomical Society, 454, 2981

\bibitem[{Dav\'{e} {et~al.}(2012)Dav\'{e}, Finlator, \&
  Oppenheimer}]{Dave_2011}
Dav\'{e}, R., Finlator, K., \& Oppenheimer, B.~D. 2012, Monthly Notices of the
  Royal Astronomical Society, 421, 98

\bibitem[{{de Blok} {et~al.}(2008){de Blok}, {Walter}, {Brinks},
  {Trachternach}, {Oh}, \& {Kennicutt}}]{deBlok_2008}
{de Blok}, W.~J.~G., {Walter}, F., {Brinks}, E., {et~al.} 2008, \aj, 136, 2648

\bibitem[{El-Badry {et~al.}(2016)El-Badry, Wetzel, Geha, Hopkins, Kereš,
  Chan, \& Faucher-GiguÚre}]{ElBadry_2016}
El-Badry, K., Wetzel, A., Geha, M., {et~al.} 2016, The Astrophysical Journal,
  820, 131

\bibitem[{El-Badry {et~al.}(2017)El-Badry, Wetzel, Geha, Quataert, Hopkins,
  Kereš, Chan, \& Faucher-GiguÚre}]{ElBadry_2017}
El-Badry, K., Wetzel, A.~R., Geha, M., {et~al.} 2017, The Astrophysical
  Journal, 835, 193

\bibitem[{El-Badry {et~al.}(2018{\natexlab{a}})El-Badry, Bradford, Quataert,
  Geha, Boylan-Kolchin, Weisz, Wetzel, Hopkins, Chan, Fitts, Kere¨, \&
  Faucher-Giguère}]{ElBadry_2018b}
El-Badry, K., Bradford, J., Quataert, E., {et~al.} 2018{\natexlab{a}}, Monthly
  Notices of the Royal Astronomical Society, 477, 1536

\bibitem[{El-Badry {et~al.}(2018{\natexlab{b}})El-Badry, Quataert, Wetzel,
  Hopkins, Weisz, Chan, Fitts, Boylan-Kolchin, Kere¨, Faucher-Giguère, \&
  Garrison-Kimmel}]{ElBadry_2018}
El-Badry, K., Quataert, E., Wetzel, A., {et~al.} 2018{\natexlab{b}}, Monthly
  Notices of the Royal Astronomical Society, 473, 1930

\bibitem[{{Emami} {et~al.}(2018){Emami}, {Siana}, {Weisz}, \&
  {Johnson}}]{Emami_2018}
{Emami}, N., {Siana}, B., {Weisz}, D.~R., \& {Johnson}, B.~D. 2018, ArXiv
  e-prints, arXiv:1809.06380

\bibitem[{Finney {et~al.}(2018)Finney, Brada{\v c}, Huang, Hoag, Morishita,
  Schrabback, Treu, Schmidt, Lemaux, Wang, \& Mason}]{Finney_2018}
Finney, E.~Q., Brada{\v c}, M., Huang, K.-H., {et~al.} 2018, The Astrophysical
  Journal, 859, 58

\bibitem[{{F{\"o}rster Schreiber} {et~al.}(2018){F{\"o}rster Schreiber},
  {Renzini}, {Mancini}, {Genzel}, {Bouch{\'e}}, {Cresci}, {Hicks}, {Lilly},
  {Peng}, {Burkert}, {Carollo}, {Cimatti}, {Daddi}, {Davies}, {Genel}, {Kurk},
  {Lang}, {Lutz}, {Mainieri}, {McCracken}, {Mignoli}, {Naab}, {Oesch},
  {Pozzetti}, {Scodeggio}, {Shapiro Griffin}, {Shapley}, {Sternberg},
  {Tacchella}, {Tacconi}, {Wuyts}, \& {Zamorani}}]{FS_2018}
{F{\"o}rster Schreiber}, N.~M., {Renzini}, A., {Mancini}, C., {et~al.} 2018,
  ArXiv e-prints, arXiv:1802.07276

\bibitem[{{Gibson} {et~al.}(2013){Gibson}, {Pilkington, K.}, {Brook, C. B.},
  {Stinson, G. S.}, \& {Bailin, J.}}]{Gibson_2013}
{Gibson}, B.~K., {Pilkington, K.}, {Brook, C. B.}, {Stinson, G. S.}, \&
  {Bailin, J.} 2013, A\&A, 554, A47

\bibitem[{Governato {et~al.}(2012)Governato, Zolotov, Pontzen, Christensen, Oh,
  Brooks, Quinn, Shen, \& Wadsley}]{Governato_2012}
Governato, F., Zolotov, A., Pontzen, A., {et~al.} 2012, Monthly Notices of the
  Royal Astronomical Society, 422, 1231

\bibitem[{Hoag {et~al.}(2017)Hoag, Brada{\v c}, Trenti, Treu, Schmidt, Huang,
  Lemaux, He, Bernard, Abramson, Mason, Morishita, Pentericci, \&
  Schrabback}]{Hoag_2017}
Hoag, A., Brada{\v c}, M., Trenti, M., {et~al.} 2017, Nature Astronomy, 1, 0091
  EP

\bibitem[{Hopkins {et~al.}(2018)Hopkins, Wetzel, Kere¨, Faucher-Giguère,
  Quataert, Boylan-Kolchin, Murray, Hayward, Garrison-Kimmel, Hummels,
  Feldmann, Torrey, Ma, Anglés-Alcázar, Su, Orr, Schmitz, Escala, Sanderson,
  Grudi?, Hafen, Kim, Fitts, Bullock, Wheeler, Chan, Elbert, \&
  Narayanan}]{Hopkins_2018}
Hopkins, P.~F., Wetzel, A., Kere¨, D., {et~al.} 2018, Monthly Notices of the
  Royal Astronomical Society, 480, 800

\bibitem[{Huang {et~al.}(2016)Huang, Lemaux, Schmidt, Hoag, Brada{\v c}, Treu,
  Dijkstra, Fontana, Henry, Malkan, Mason, Morishita, Pentericci, Jr., Trenti,
  \& Wang}]{Huang_2016}
Huang, K.-H., Lemaux, B.~C., Schmidt, K.~B., {et~al.} 2016, The Astrophysical
  Journal Letters, 823, L14

\bibitem[{{Hung} {et~al.}(2018){Hung}, {Hayward}, {Yuan}, {Boylan-Kolchin},
  {Faucher-Gigu{\`e}re}, {Hopkins}, {Kere{\v s}}, {Murray}, \&
  {Wetzel}}]{Hung_2018}
{Hung}, C.-L., {Hayward}, C.~C., {Yuan}, T., {et~al.} 2018, ArXiv e-prints,
  arXiv:1806.04233

\bibitem[{{Jones} {et~al.}(2010){Jones}, {Ellis}, {Jullo}, \&
  {Richard}}]{Jones_2010}
{Jones}, T., {Ellis}, R., {Jullo}, E., \& {Richard}, J. 2010, \apj, 725, L176

\bibitem[{Jones {et~al.}(2015)Jones, Wang, Schmidt, Treu, Brammer, Brada{\v
  c}, Dressler, Henry, Malkan, Pentericci, \& Trenti}]{Jones_2015}
Jones, T., Wang, X., Schmidt, K.~B., {et~al.} 2015, The Astronomical Journal,
  149, 107

\bibitem[{Kennicutt(1998)}]{Kennicutt_1998}
Kennicutt, R.~C. 1998, Annual Review of Astronomy and Astrophysics, 36, 189

\bibitem[{Kriek {et~al.}(2009)Kriek, van Dokkum, Labbé, Franx, Illingworth,
  Marchesini, \& Quadri}]{Kriek_2009}
Kriek, M., van Dokkum, P.~G., Labbé, I., {et~al.} 2009, The Astrophysical
  Journal, 700, 221

\bibitem[{{Larkin} {et~al.}(2006){Larkin}, {Barczys}, {Krabbe}, {Adkins},
  {Aliado}, {Amico}, {Brims}, {Campbell}, {Canfield}, {Gasaway}, {Honey},
  {Iserlohe}, {Johnson}, {Kress}, {LaFreniere}, {Lyke}, {Magnone}, {Magnone},
  {McElwain}, {Moon}, {Quirrenbach}, {Skulason}, {Song}, {Spencer}, {Weiss}, \&
  {Wright}}]{Larkin_2006}
{Larkin}, J., {Barczys}, M., {Krabbe}, A., {et~al.} 2006, in \procspie, Vol.
  6269, Society of Photo-Optical Instrumentation Engineers (SPIE) Conference
  Series, 62691A

\bibitem[{Leethochawalit {et~al.}(2016)Leethochawalit, Jones, Ellis, Stark,
  Richard, Zitrin, \& Auger}]{Leethochawalit_2016}
Leethochawalit, N., Jones, T.~A., Ellis, R.~S., {et~al.} 2016, The
  Astrophysical Journal, 820, 84

\bibitem[{{Livermore} {et~al.}(2015){Livermore}, {Jones}, {Richard}, {Bower},
  {Swinbank}, {Yuan}, {Edge}, {Ellis}, {Kewley}, {Smail}, {Coppin}, \&
  {Ebeling}}]{Livermore_2015}
{Livermore}, R.~C., {Jones}, T.~A., {Richard}, J., {et~al.} 2015, \mnras, 450,
  1812

\bibitem[{Lotz {et~al.}(2017)Lotz, Koekemoer, Coe, Grogin, Capak, Mack,
  Anderson, Avila, Barker, Borncamp, Brammer, Durbin, Gunning, Hilbert,
  Jenkner, Khandrika, Levay, Lucas, MacKenty, Ogaz, Porterfield, Reid,
  Robberto, Royle, Smith, Storrie-Lombardi, Sunnquist, Surace, Taylor,
  Williams, Bullock, Dickinson, Finkelstein, Natarajan, Richard, Robertson,
  Tumlinson, Zitrin, Flanagan, Sembach, Soifer, \& Mountain}]{Lotz_2017}
Lotz, J.~M., Koekemoer, A., Coe, D., {et~al.} 2017, The Astrophysical Journal,
  837, 97

\bibitem[{Ma {et~al.}(2017)Ma, Hopkins, Feldmann, Torrey, Faucher-Giguère, \&
  Kere¨}]{Ma_2017}
Ma, X., Hopkins, P.~F., Feldmann, R., {et~al.} 2017, Monthly Notices of the
  Royal Astronomical Society, 466, 4780

\bibitem[{Madau \& Dickinson(2014)}]{Madau_Dickinson_2014}
Madau, P., \& Dickinson, M. 2014, Annual Review of Astronomy and Astrophysics,
  52, 415

\bibitem[{{Mason} {et~al.}(2017){Mason}, {Treu}, {Fontana}, {Jones},
  {Morishita}, {Amorin}, {Brada{\v{c}}}, {Quinn Finney}, {Grillo}, {Henry},
  {Hoag}, {Huang}, {Schmidt}, {Trenti}, \& {Vulcani}}]{Mason_2017}
{Mason}, C.~A., {Treu}, T., {Fontana}, A., {et~al.} 2017, \apj, 838, 14

\bibitem[{{McGaugh} {et~al.}(2001){McGaugh}, {Rubin}, \& {de
  Blok}}]{McGaugh_2001}
{McGaugh}, S.~S., {Rubin}, V.~C., \& {de Blok}, W.~J.~G. 2001, \aj, 122, 2381

\bibitem[{Meurer {et~al.}(1999)Meurer, Heckman, \& Calzetti}]{Meurer_1999}
Meurer, G.~R., Heckman, T.~M., \& Calzetti, D. 1999, The Astrophysical Journal,
  521, 64

\bibitem[{{Navarro} {et~al.}(1996){Navarro}, {Frenk}, \&
  {White}}]{Navarro_1996}
{Navarro}, J.~F., {Frenk}, C.~S., \& {White}, S.~D.~M. 1996, \apj, 462, 563

\bibitem[{{Navarro} {et~al.}(2010){Navarro}, {Ludlow}, {Springel}, {Wang},
  {Vogelsberger}, {White}, {Jenkins}, {Frenk}, \& {Helmi}}]{Navarro_2010}
{Navarro}, J.~F., {Ludlow}, A., {Springel}, V., {et~al.} 2010, \mnras, 402, 21

\bibitem[{{O{\~n}orbe} {et~al.}(2015){O{\~n}orbe}, {Boylan-Kolchin}, {Bullock},
  {Hopkins}, {Kere{\v s}}, {Faucher-Gigu{\`e}re}, {Quataert}, \&
  {Murray}}]{Onorbe_2015}
{O{\~n}orbe}, J., {Boylan-Kolchin}, M., {Bullock}, J.~S., {et~al.} 2015,
  \mnras, 454, 2092

\bibitem[{{Pontzen} \& {Governato}(2012)}]{Pontzen_2012}
{Pontzen}, A., \& {Governato}, F. 2012, \mnras, 421, 3464

\bibitem[{Postman {et~al.}(2012)Postman, Coe, Benítez, Bradley, Broadhurst,
  Donahue, Ford, Graur, Graves, Jouvel, Koekemoer, Lemze, Medezinski, Molino,
  Moustakas, Ogaz, Riess, Rodney, Rosati, Umetsu, Zheng, Zitrin, Bartelmann,
  Bouwens, Czakon, Golwala, Host, Infante, Jha, Jimenez-Teja, Kelson, Lahav,
  Lazkoz, Maoz, McCully, Melchior, Meneghetti, Merten, Moustakas, Nonino,
  Patel, Regös, Sayers, Seitz, \& der Wel}]{Postman_2012}
Postman, M., Coe, D., Benítez, N., {et~al.} 2012, The Astrophysical Journal
  Supplement Series, 199, 25

\bibitem[{Read {et~al.}(2016)Read, Agertz, \& Collins}]{Read_2016}
Read, J.~I., Agertz, O., \& Collins, M. L.~M. 2016, Monthly Notices of the
  Royal Astronomical Society, 459, 2573

\bibitem[{{Read} {et~al.}(2018){Read}, {Walker}, \& {Steger}}]{Read_2018}
{Read}, J.~I., {Walker}, M.~G., \& {Steger}, P. 2018, ArXiv e-prints,
  arXiv:1808.06634

\bibitem[{Simons {et~al.}(2016)Simons, Kassin, Trump, Weiner, Heckman, Barro,
  Koo, Guo, Pacifici, Koekemoer, \& Stephens}]{Simons_2016}
Simons, R.~C., Kassin, S.~A., Trump, J.~R., {et~al.} 2016, The Astrophysical
  Journal, 830, 14

\bibitem[{{Sparre} {et~al.}(2017){Sparre}, {Hayward}, {Feldmann},
  {Faucher-Gigu{\`e}re}, {Muratov}, {Kere{\v s}}, \& {Hopkins}}]{Sparre_2016}
{Sparre}, M., {Hayward}, C.~C., {Feldmann}, R., {et~al.} 2017, \mnras, 466, 88

\bibitem[{Spergel \& Steinhardt(2000)}]{Spergel_2000}
Spergel, D.~N., \& Steinhardt, P.~J. 2000, Phys. Rev. Lett., 84, 3760

\bibitem[{{Stark} {et~al.}(2008){Stark}, {Swinbank}, {Ellis}, {Dye}, {Smail},
  \& {Richard}}]{Stark_2008}
{Stark}, D.~P., {Swinbank}, A.~M., {Ellis}, R.~S., {et~al.} 2008, \nat, 455,
  775

\bibitem[{{Stott} {et~al.}(2016){Stott}, {Swinbank}, {Johnson}, {Tiley},
  {Magdis}, {Bower}, {Bunker}, {Bureau}, {Harrison}, {Jarvis}, {Sharples},
  {Smail}, {Sobral}, {Best}, \& {Cirasuolo}}]{Stott_2016}
{Stott}, J.~P., {Swinbank}, A.~M., {Johnson}, H.~L., {et~al.} 2016, \mnras,
  457, 1888

\bibitem[{{Strait} {et~al.}(2018){Strait}, {Brada{\v c}}, {Hoag}, {Huang},
  {Treu}, {Wang}, {Amorin}, {Castellano}, {Fontana}, {Lemaux}, {Merlin},
  {Schmidt}, {Schrabback}, {Tomczack}, {Trenti}, \& {Vulcani}}]{Strait_2018}
{Strait}, V., {Brada{\v c}}, M., {Hoag}, A., {et~al.} 2018, ArXiv e-prints,
  arXiv:1805.08789

\bibitem[{Teyssier {et~al.}(2013)Teyssier, Pontzen, Dubois, \&
  Read}]{Teyssier_2013}
Teyssier, R., Pontzen, A., Dubois, Y., \& Read, J.~I. 2013, Monthly Notices of
  the Royal Astronomical Society, 429, 3068

\bibitem[{{Treu} {et~al.}(2015){Treu}, {Schmidt}, {Brammer}, {Vulcani}, {Wang},
  {Brada{\v c}}, {Dijkstra}, {Dressler}, {Fontana}, {Gavazzi}, {Henry}, {Hoag},
  {Huang}, {Jones}, {Kelly}, {Malkan}, {Mason}, {Pentericci}, {Poggianti},
  {Stiavelli}, {Trenti}, \& {von der Linden}}]{Treu_2015}
{Treu}, T., {Schmidt}, K.~B., {Brammer}, G.~B., {et~al.} 2015, \apj, 812, 114

\bibitem[{Viel {et~al.}(2005)Viel, Lesgourgues, Haehnelt, Matarrese, \&
  Riotto}]{Viel_2005}
Viel, M., Lesgourgues, J., Haehnelt, M.~G., Matarrese, S., \& Riotto, A. 2005,
  Phys. Rev. D, 71, 063534

\bibitem[{Vogelsberger {et~al.}(2014)Vogelsberger, Zavala, Simpson, \&
  Jenkins}]{Vogelsberger_2014}
Vogelsberger, M., Zavala, J., Simpson, C., \& Jenkins, A. 2014, Monthly Notices
  of the Royal Astronomical Society, 444, 3684

\bibitem[{Wang {et~al.}(2017)Wang, Jones, Treu, Morishita, Abramson, Brammer,
  Huang, Malkan, Schmidt, Fontana, Grillo, Henry, Karman, Kelly, Mason,
  Mercurio, Rosati, Sharon, Trenti, \& Vulcani}]{Wang_2017}
Wang, X., Jones, T.~A., Treu, T., {et~al.} 2017, The Astrophysical Journal,
  837, 89

\bibitem[{{Wang} {et~al.}(2018){Wang}, {Jones}, {Treu}, {Hirtenstein},
  {Brammer}, {Daddi}, {Meng}, {Morishita}, {Abramson}, {Henry}, {Peng},
  {Schmidt}, {Sharon}, {Trenti}, \& {Vulcani}}]{Wang_2018}
{Wang}, X., {Jones}, T.~A., {Treu}, T., {et~al.} 2018, ArXiv e-prints,
  arXiv:1808.08800

\bibitem[{Wuyts {et~al.}(2016)Wuyts, Schreiber, Wisnioski, Genzel, Burkert,
  Bandara, Beifiori, Belli, Bender, Brammer, Chan, Davies, Fossati, Galametz,
  Kulkarni, Lang, Lutz, Mendel, Momcheva, Naab, Nelson, Saglia, Seitz, Tacconi,
  ichi Tadaki, Übler, van Dokkum, Wilman, \& Wuyts}]{Wuyts_2016}
Wuyts, S., Schreiber, N. M.~F., Wisnioski, E., {et~al.} 2016, The Astrophysical
  Journal, 831, 149

\bibitem[{Zavala {et~al.}(2013)Zavala, Vogelsberger, \& Walker}]{Zavala_2013}
Zavala, J., Vogelsberger, M., \& Walker, M.~G. 2013, Monthly Notices of the
  Royal Astronomical Society: Letters, 431, L20

\end{thebibliography}




%
%
%

\begin{appendix}
\section{FIRE Simulations}
\label{app:sims}

Table \ref{tab:sims} presents properties of the simulated galaxies from FIRE that were used in this analysis, as described in Section \ref{sec:sims}.

\begin{table*}[h]
\centering
\caption{Simulated Galaxies}
\begin{tabular}{c c c c c c c}
\hline
galaxy & resolution & mass (z=1.5)\footnote{z $\sim$ 1.5 ranges from z = 1.32 - 1.72} & $\sigma$ (z=1.5) & mass (z=2)\footnote{z $\sim$ 2 ranges from z = 1.89 - 2.57} & $\sigma$ (z=2) & citation\\
 & & $\log(M_*/M_{\odot})$ & km/s & $\log(M_*/M_{\odot})$& km/s &\\
 \hline
m11i & 7100 & 7.64 & 6-38 & - & - &\cite{ElBadry_2018b}\\
m11q & 880 & 8.05 & 4-66 & - & - & \cite{Hopkins_2018}\\
m11d & 7100 & 8.39 & 5-43 & 8.25 & 9-40 & \cite{ElBadry_2018b}\\
m11e & 7100 & 8.37& 3-70 & 7.82 & 7- 44& \cite{ElBadry_2018b}\\
m11h & 7100 & 8.67 & 15-52 & 8.38 & 12-68 & \cite{ElBadry_2018b}\\
m12z & 4200 & 8.88 & 15-75 & 8.53 & 11-48 & \cite{Hopkins_2018}\\
m12c & 7100 & 9.23 & 13-54 & - & - & \cite{Hopkins_2018}\\
m12m & 7100 & 9.68 & 31-125 & 9.29 & 12-92 & \cite{Hopkins_2018}\\
m12i & 7100 & 9.74& 28-80 & 9.36 & 20-89 & \cite{Hopkins_2018}\\
\end{tabular}%
\label{tab:sims}
\end{table*}

\section{Kinematic maps}
\label{app:kin_maps}

\begin{figure*}[h]
\centering
  \includegraphics[width=0.9\linewidth]{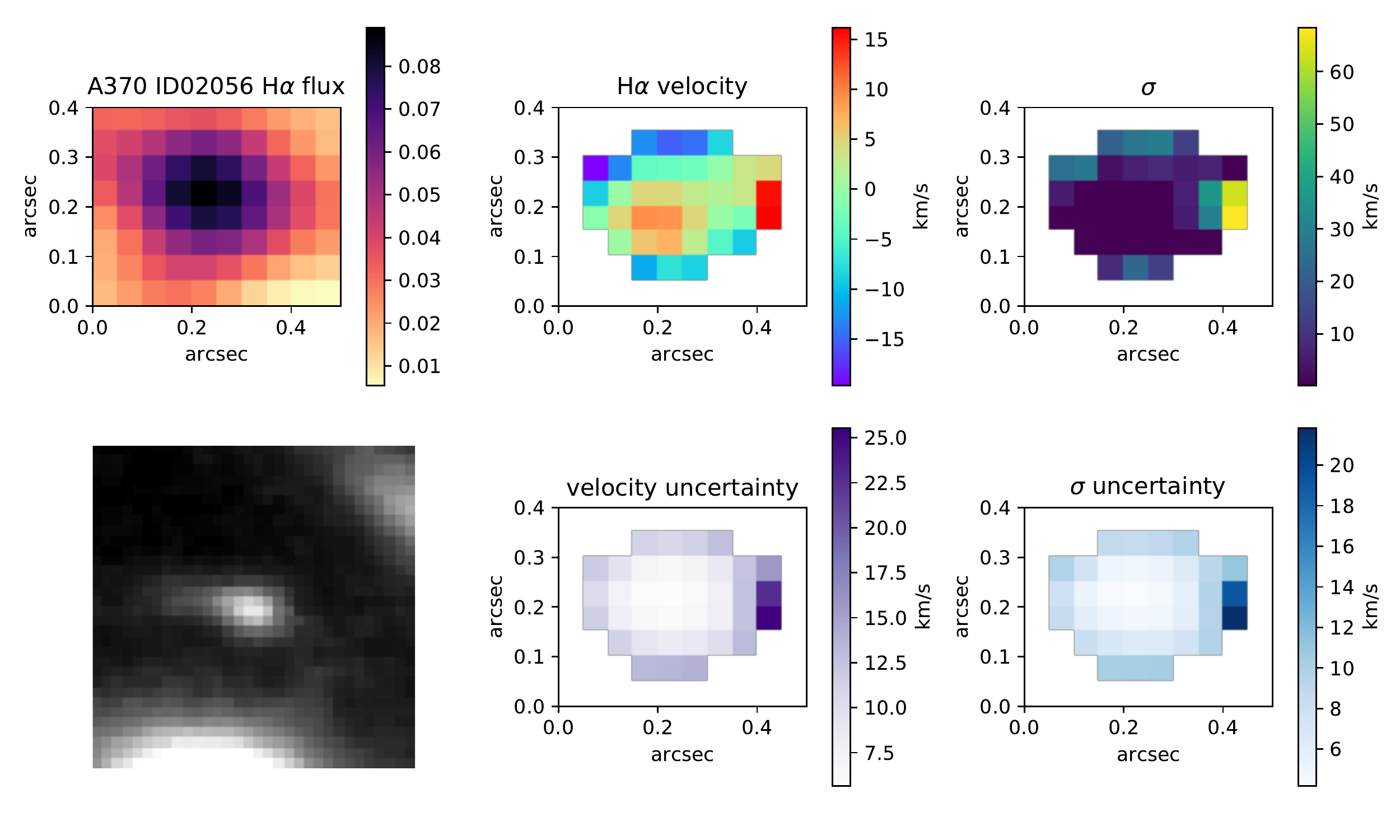}
\caption{From left to right: (Top) H$\alpha$ flux map, H$\alpha$ velocity map, $\sigma$ map, (Bottom) Hubble Image, velocity uncertainty map, $\sigma$ uncertainty map for A370 ID02056. The HST image field of view is 2x2 arcseconds.}
\label{fig:appendix}
\end{figure*}

\begin{figure*}
\centering
  \includegraphics[width=0.9\linewidth]{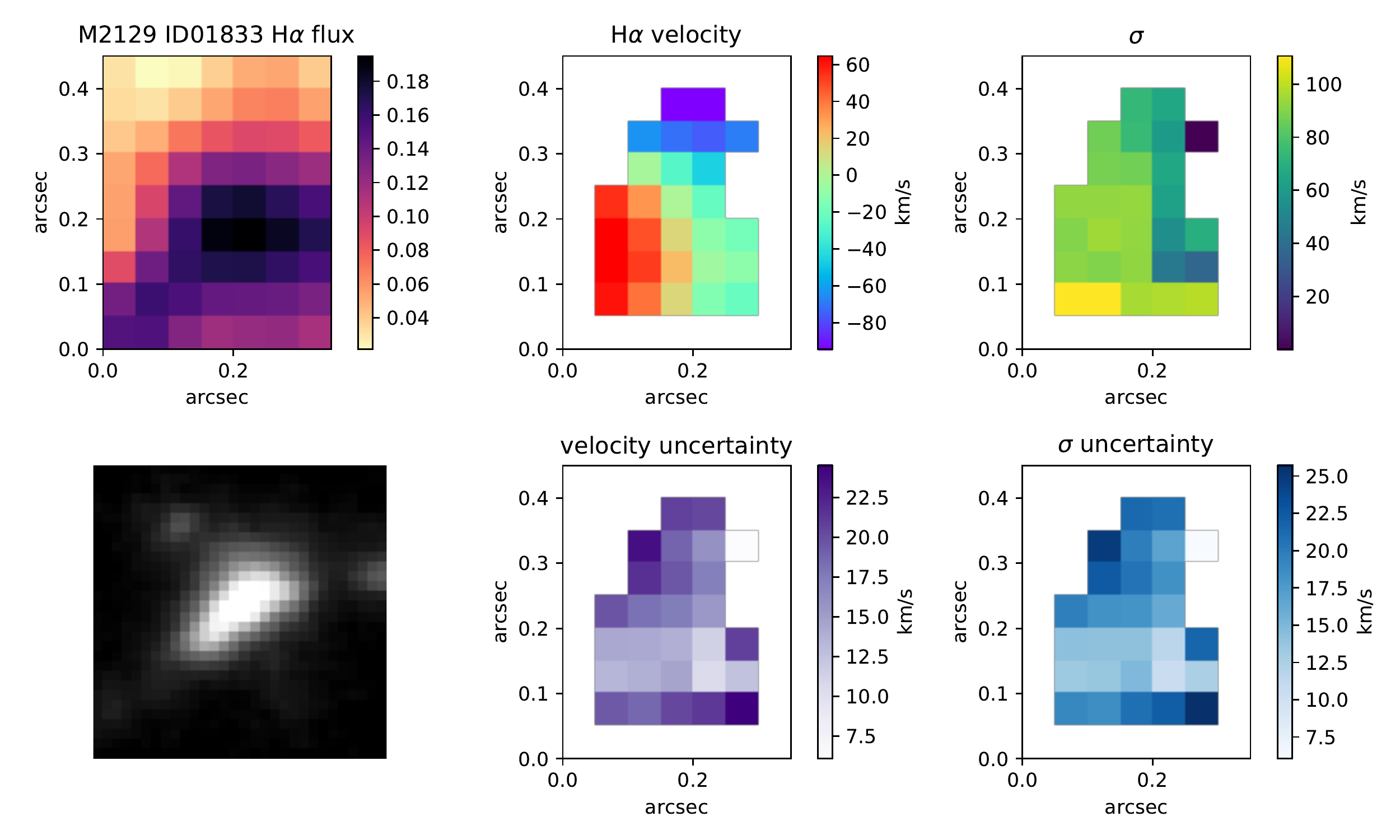}
  \caption{Same as Figure \ref{fig:appendix}, for M2129 ID01833}
\label{fig:applast}
\end{figure*}

Maps of kinematics, H$\alpha$ emission, and HST broad-band images of each target are presented in Figures \ref{fig:appendix} - \ref{fig:applast}. The methods for generating these maps are described in Section \ref{sec:methods}.

\begin{figure*}
\centering
  \includegraphics[width=0.9\linewidth]{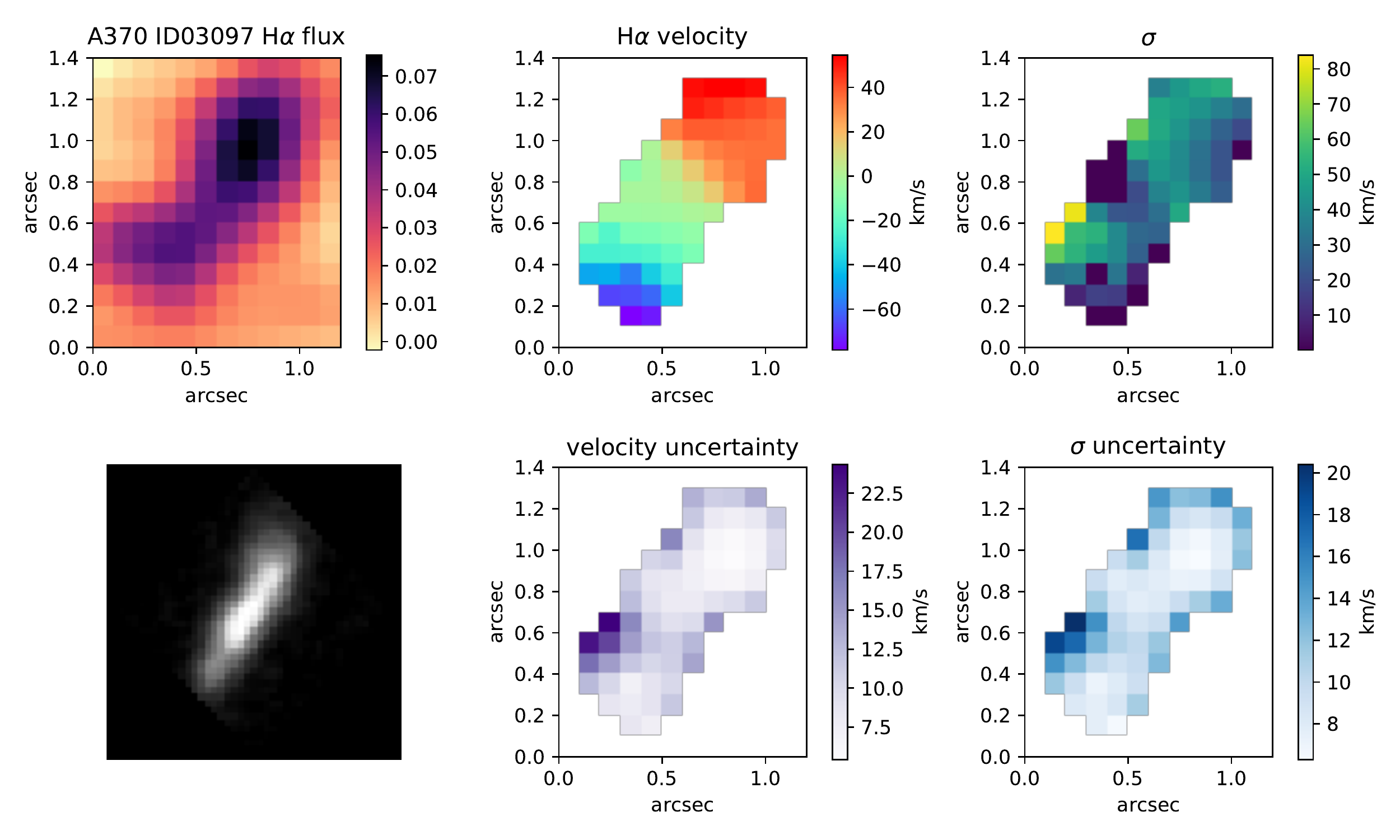}
\caption{Same as Figure \ref{fig:appendix}, for A370 ID03097}
\end{figure*}

\begin{figure*}
\centering
  \includegraphics[width=0.9\linewidth]{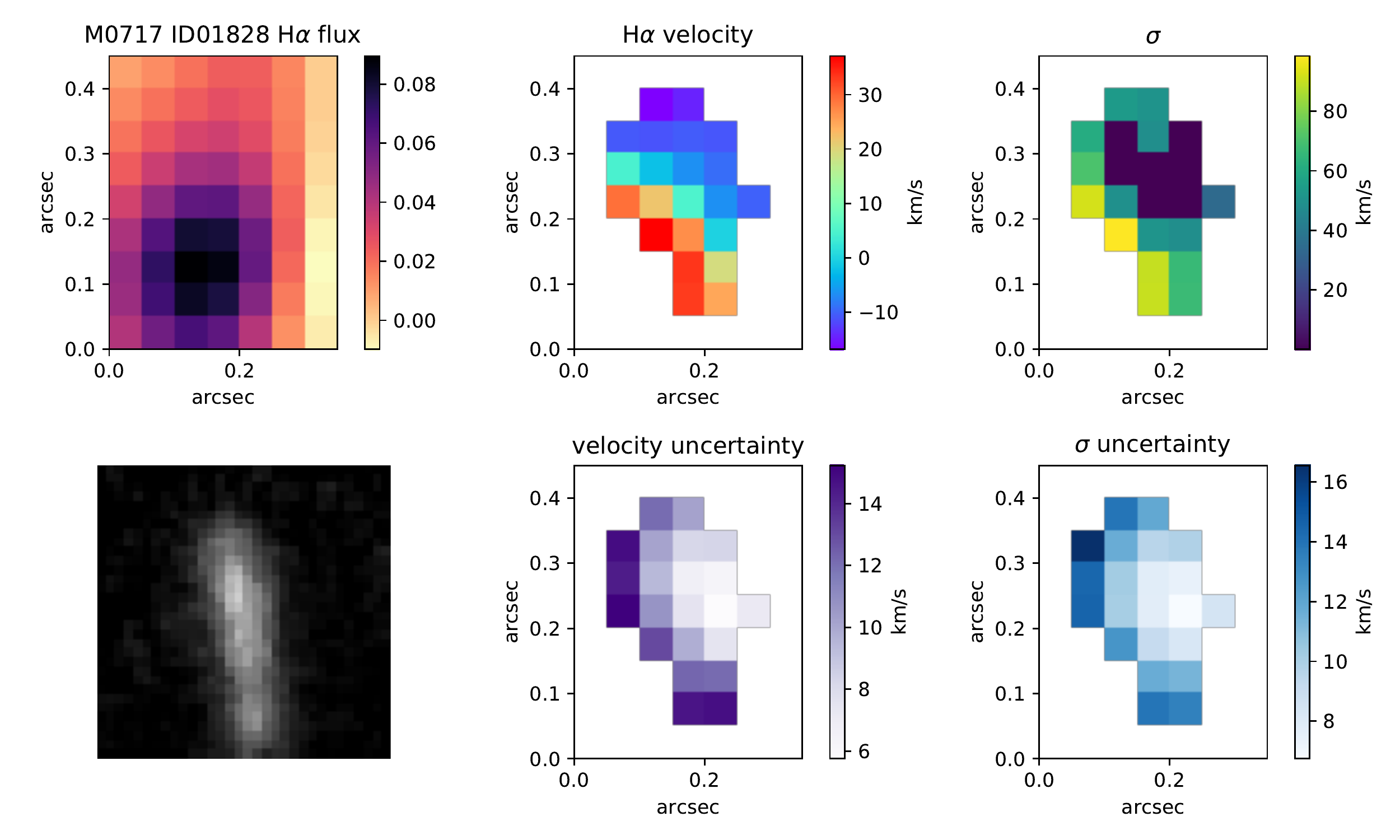}
  \caption{Same as Figure \ref{fig:appendix}, for M0717 ID01828}
\end{figure*}

\begin{figure*}
\centering
  \includegraphics[width=0.9\linewidth]{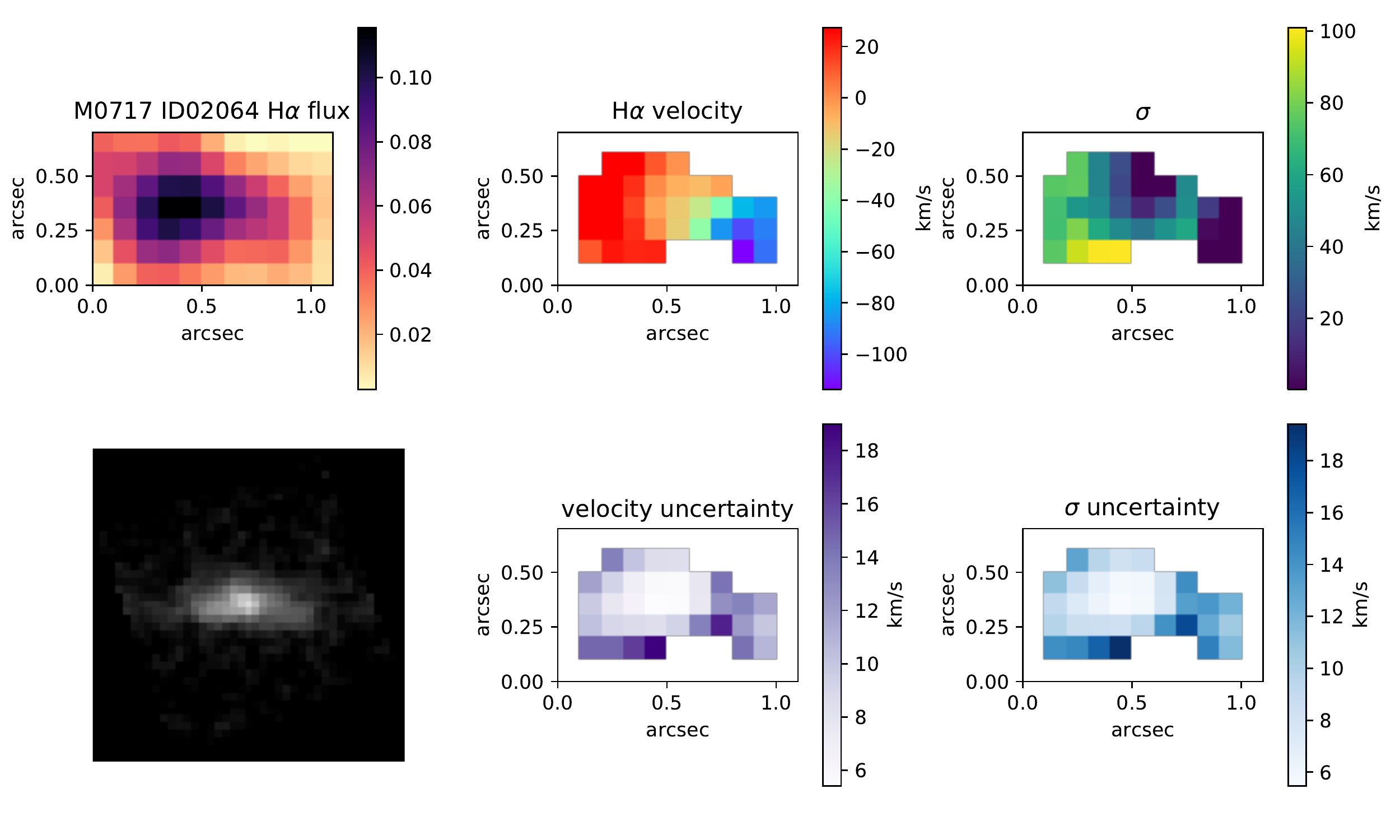}
  \caption{Same as Figure \ref{fig:appendix}, for M0717 ID02064}
\end{figure*}

\begin{figure*}
\centering
  \includegraphics[width=0.9\linewidth]{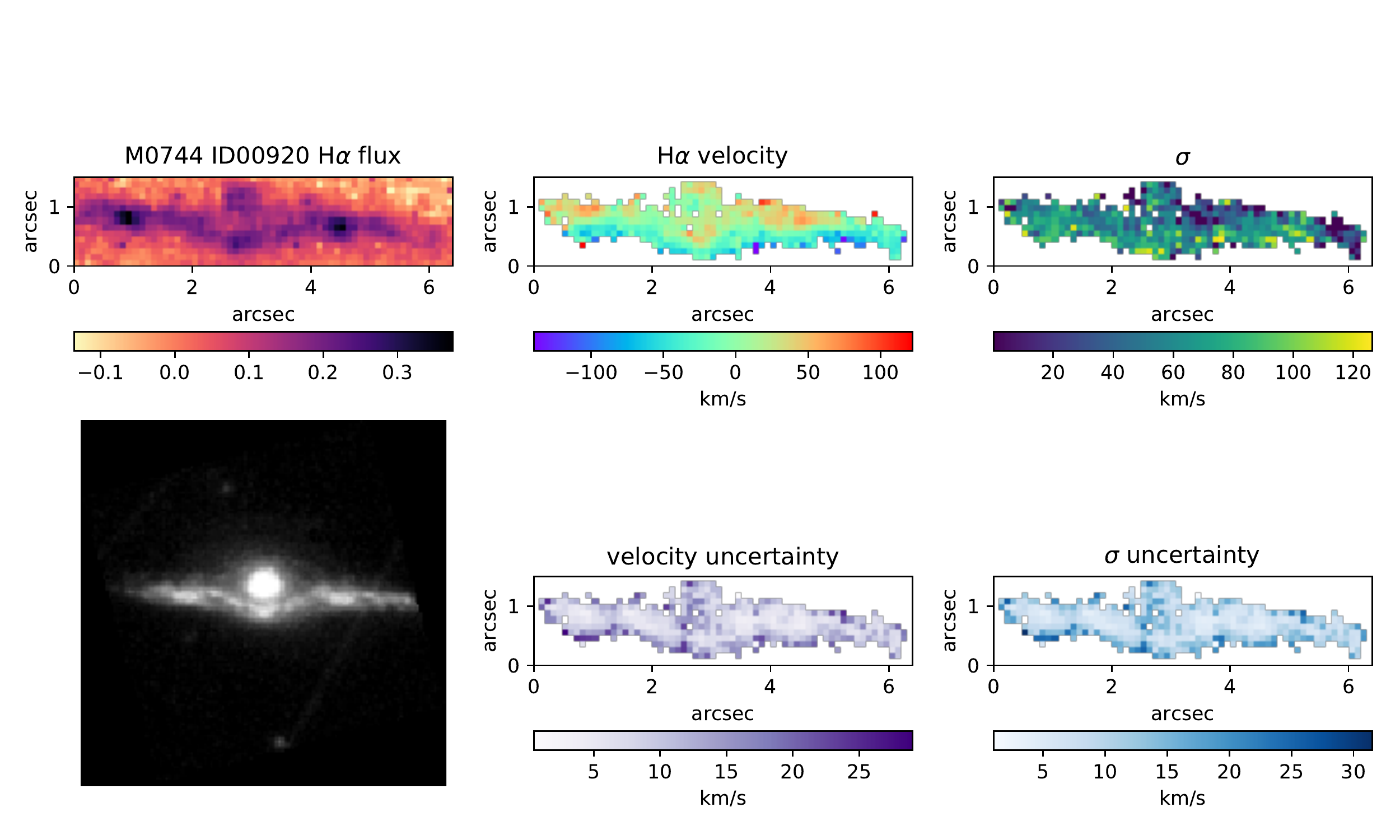}
  \caption{Same as Figure \ref{fig:appendix}, for M0744 ID00920. The HST image field of view is 7x7 arcseconds.}
\end{figure*}

\begin{figure*}
\centering
  \includegraphics[width=0.9\linewidth]{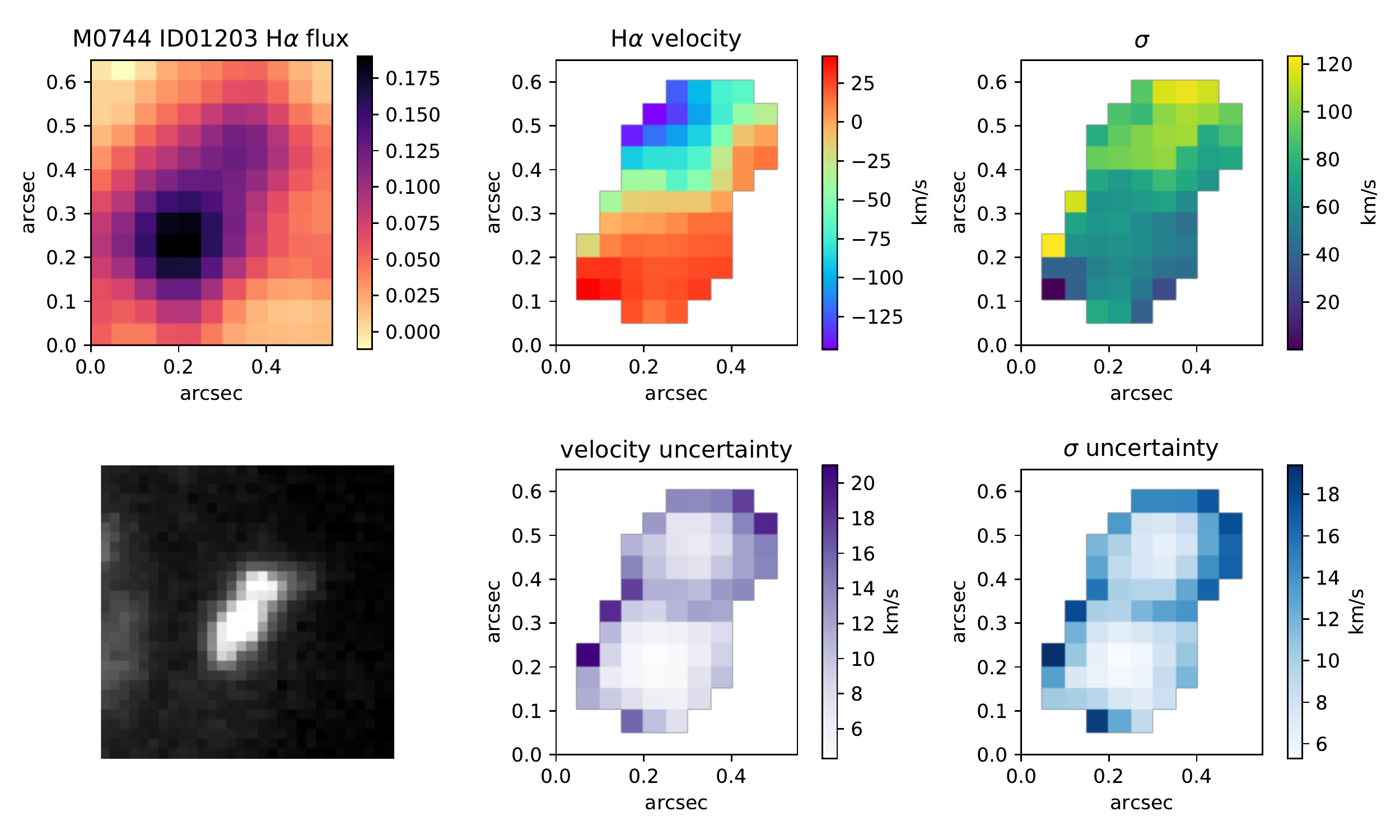}
  \caption{Same as Figure \ref{fig:appendix}, for M0744 ID01203}
\end{figure*}

\begin{figure*}
\centering
  \includegraphics[width=0.9\linewidth]{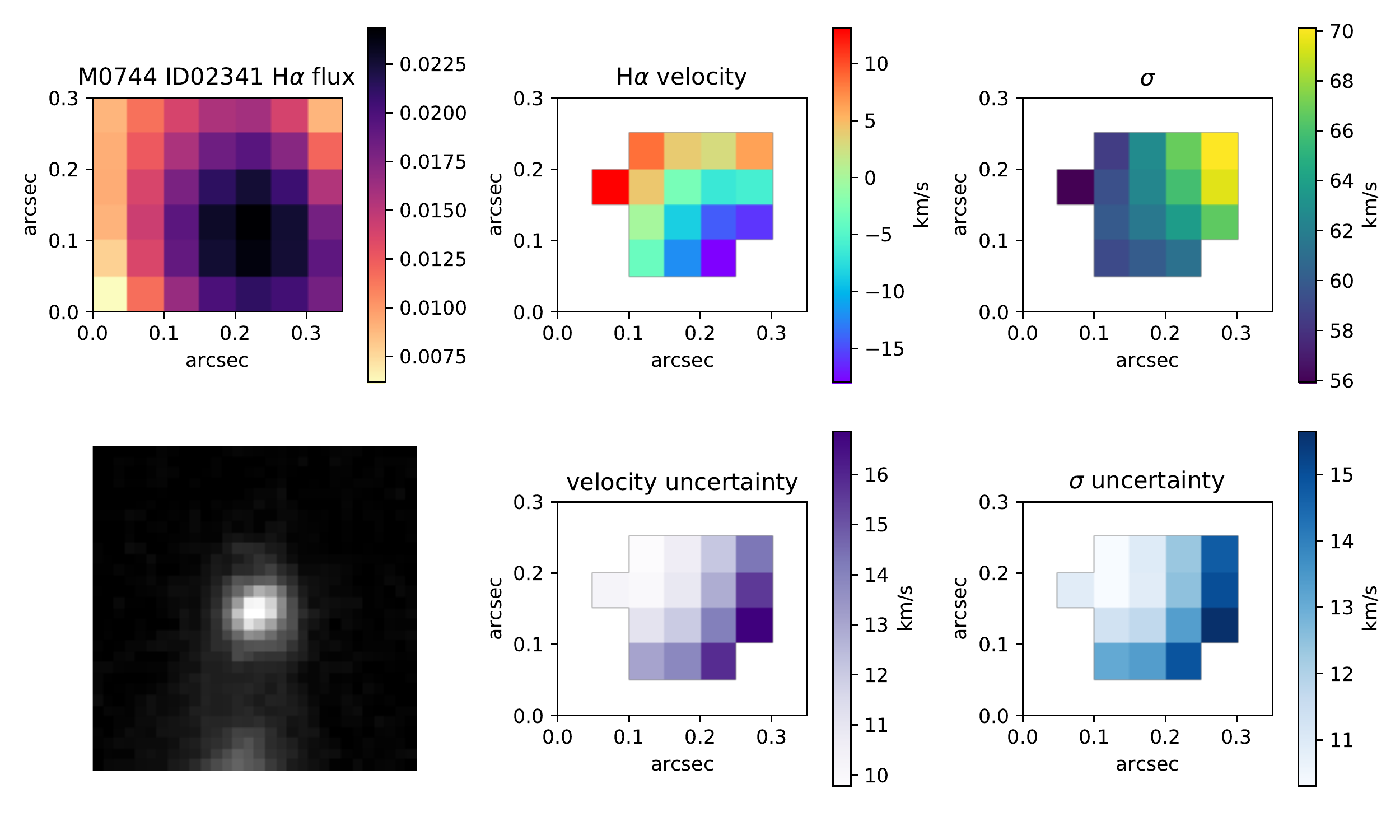}
  \caption{Same as Figure \ref{fig:appendix}, for M0744 ID02341}
\end{figure*}

\begin{figure*}
\centering
  \includegraphics[width=0.9\linewidth]{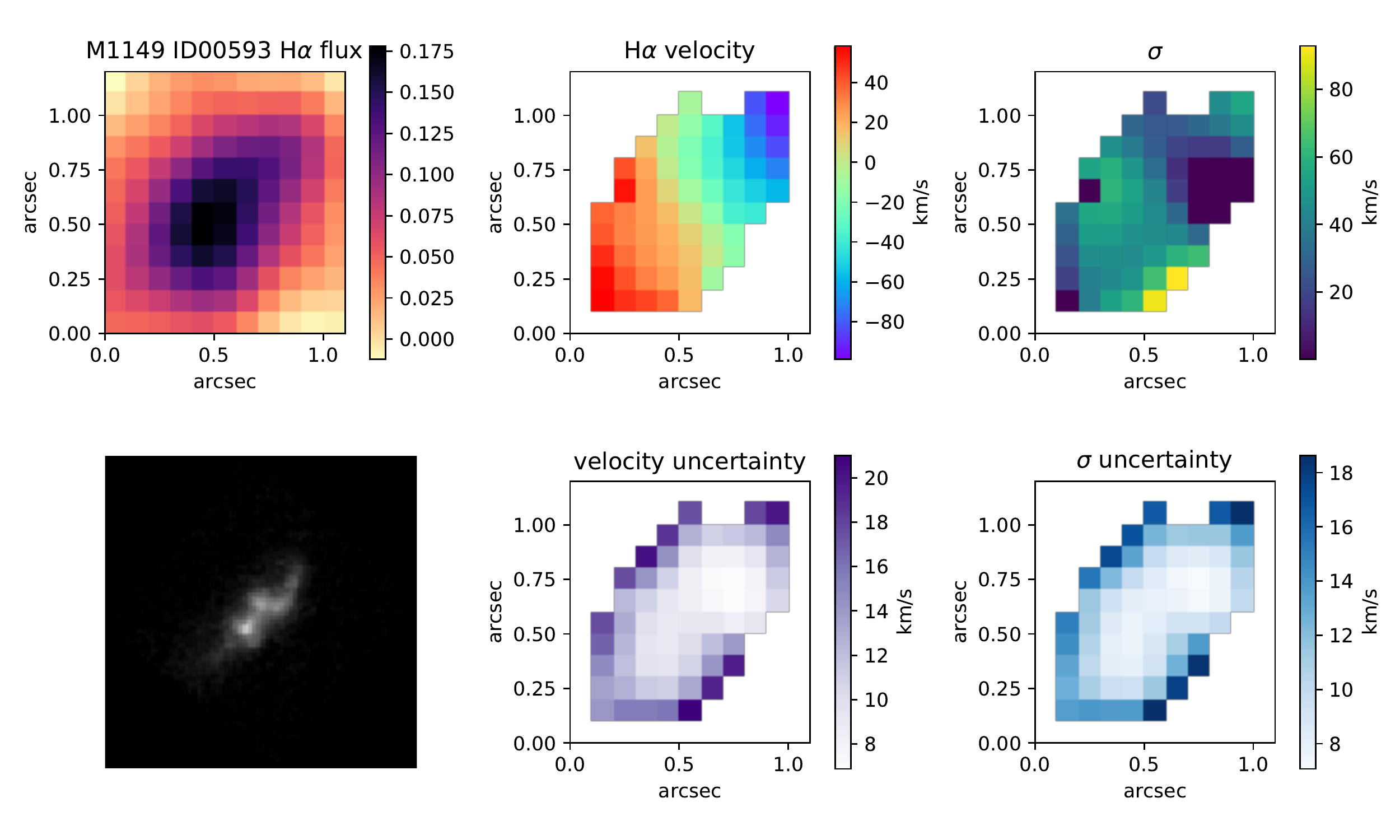}
  \caption{Same as Figure \ref{fig:appendix}, for M1149 ID00593}
\end{figure*}

\begin{figure*}
\centering
  \includegraphics[width=0.9\linewidth]{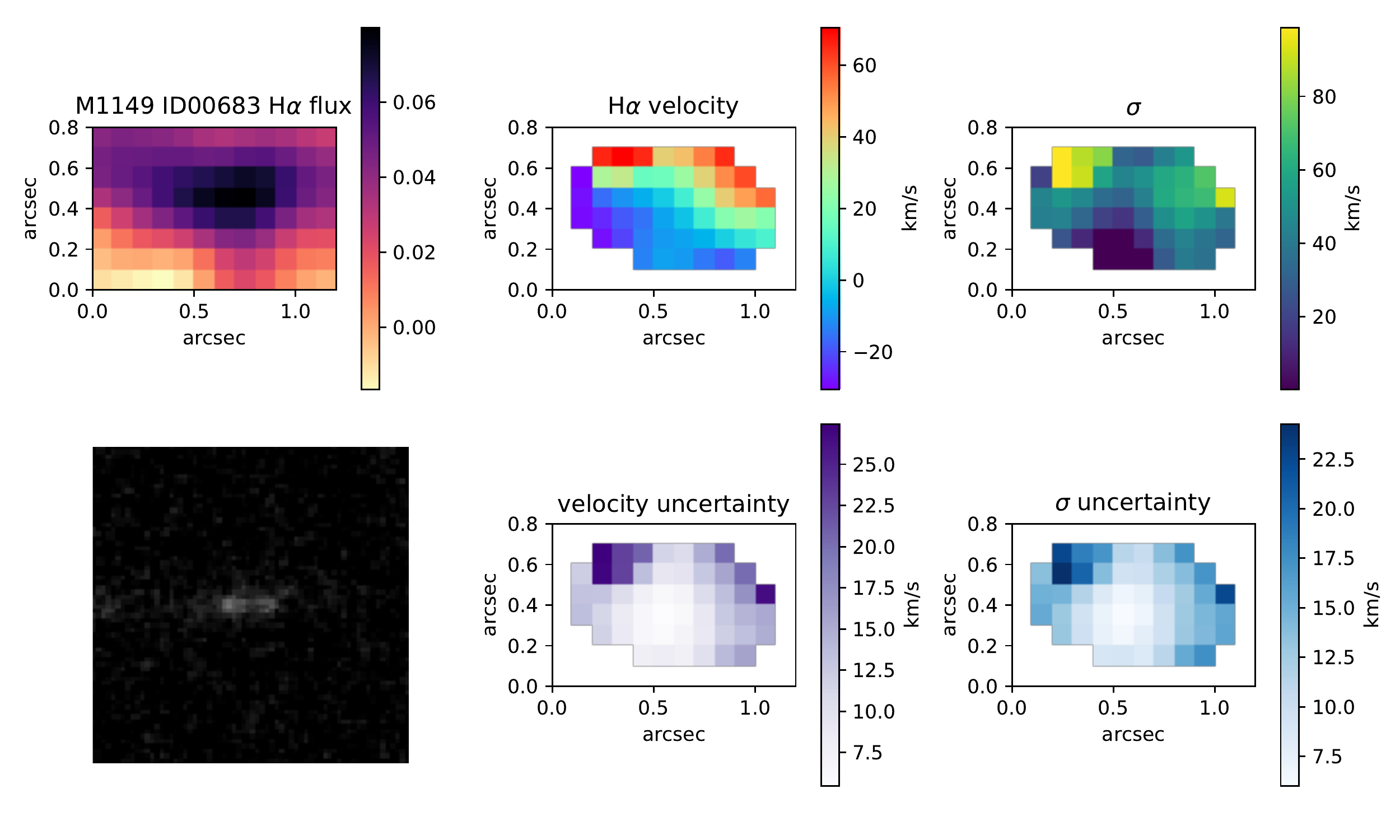}
  \caption{Same as Figure \ref{fig:appendix}, for M1149 ID00683}
\end{figure*}

\begin{figure*}
\centering
  \includegraphics[width=0.9\linewidth]{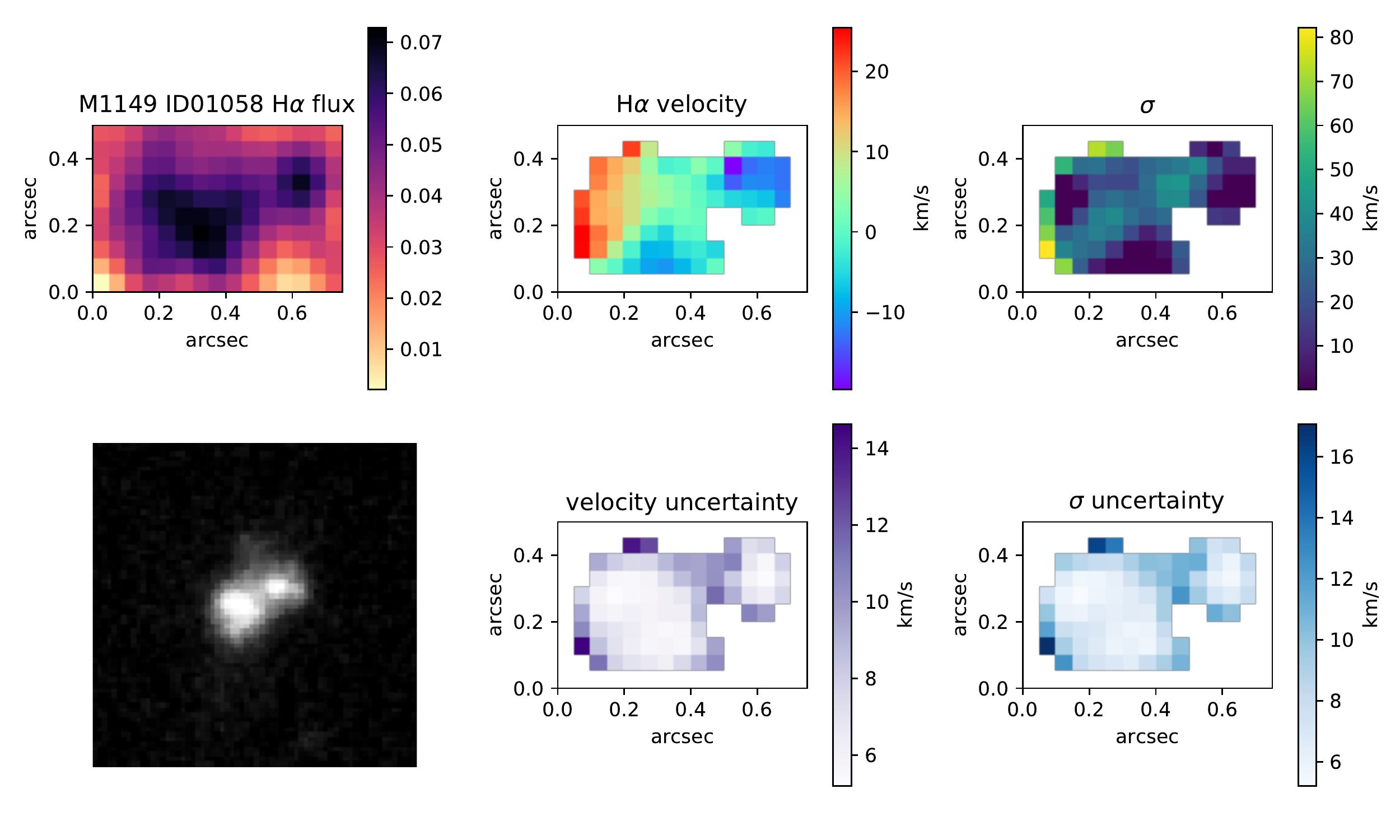}
  \caption{Same as Figure \ref{fig:appendix}, for M1149 ID01058}
\end{figure*}

\begin{figure*}
\centering
  \includegraphics[width=0.9\linewidth]{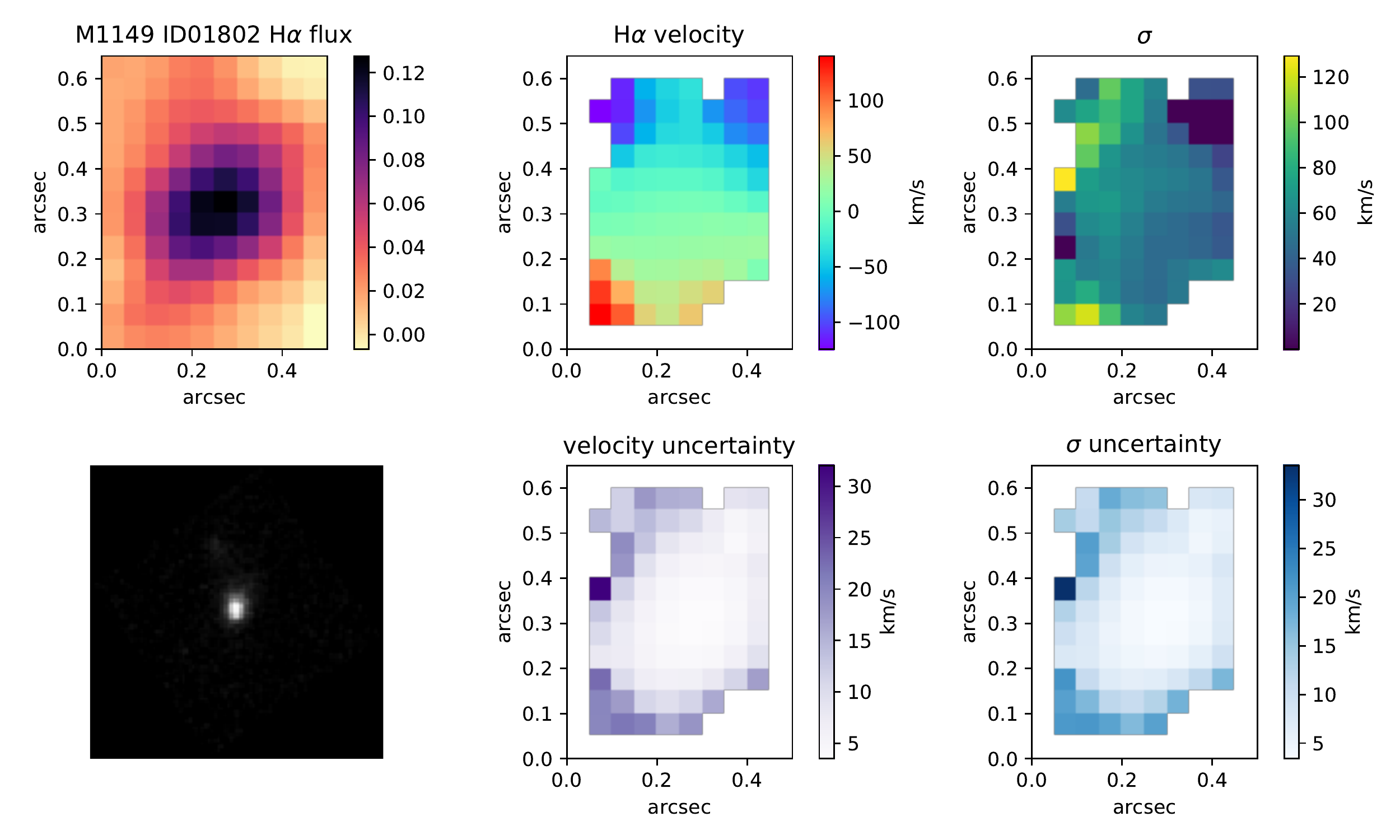}
  \caption{Same as Figure \ref{fig:appendix}, for M1149 ID01802}
\end{figure*}

\begin{figure*}
\centering
  \includegraphics[width=0.9\linewidth]{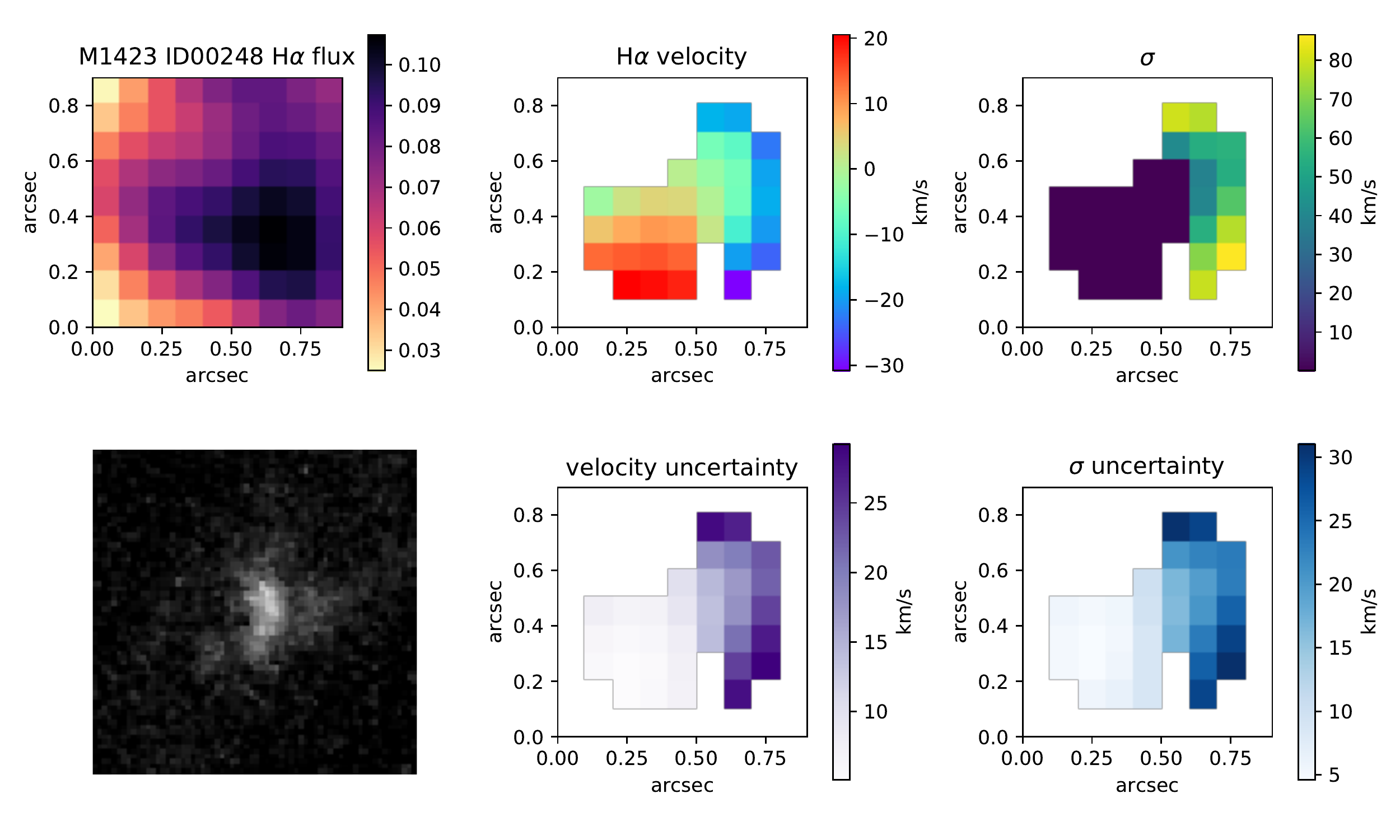}
  \caption{Same as Figure \ref{fig:appendix}, for M1423 ID00248}
\end{figure*}

\begin{figure*}
\centering
  \includegraphics[width=0.9\linewidth]{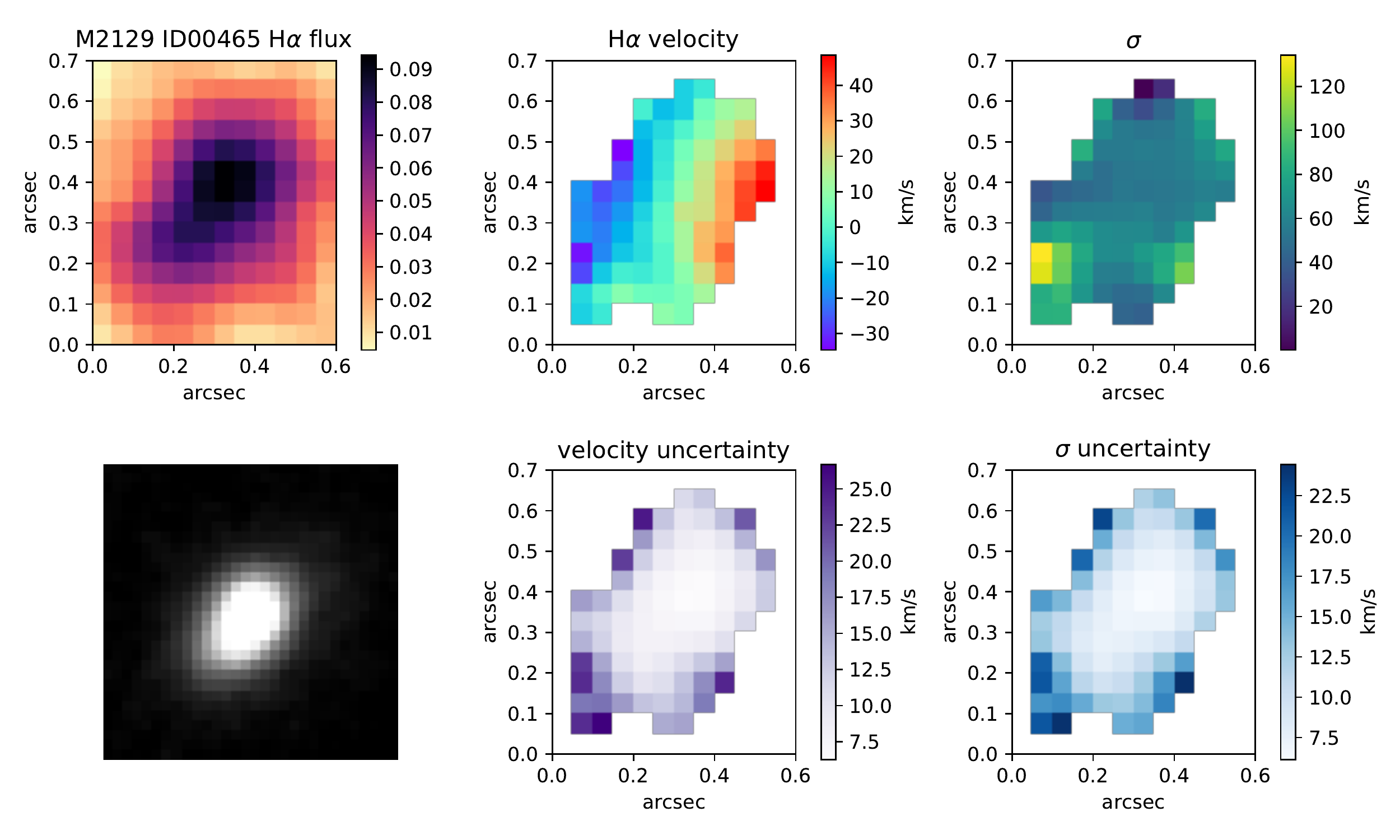}
  \caption{Same as Figure \ref{fig:appendix}, for M2129 ID00465}
\end{figure*}

\begin{figure*}
\centering
  \includegraphics[width=0.9\linewidth]{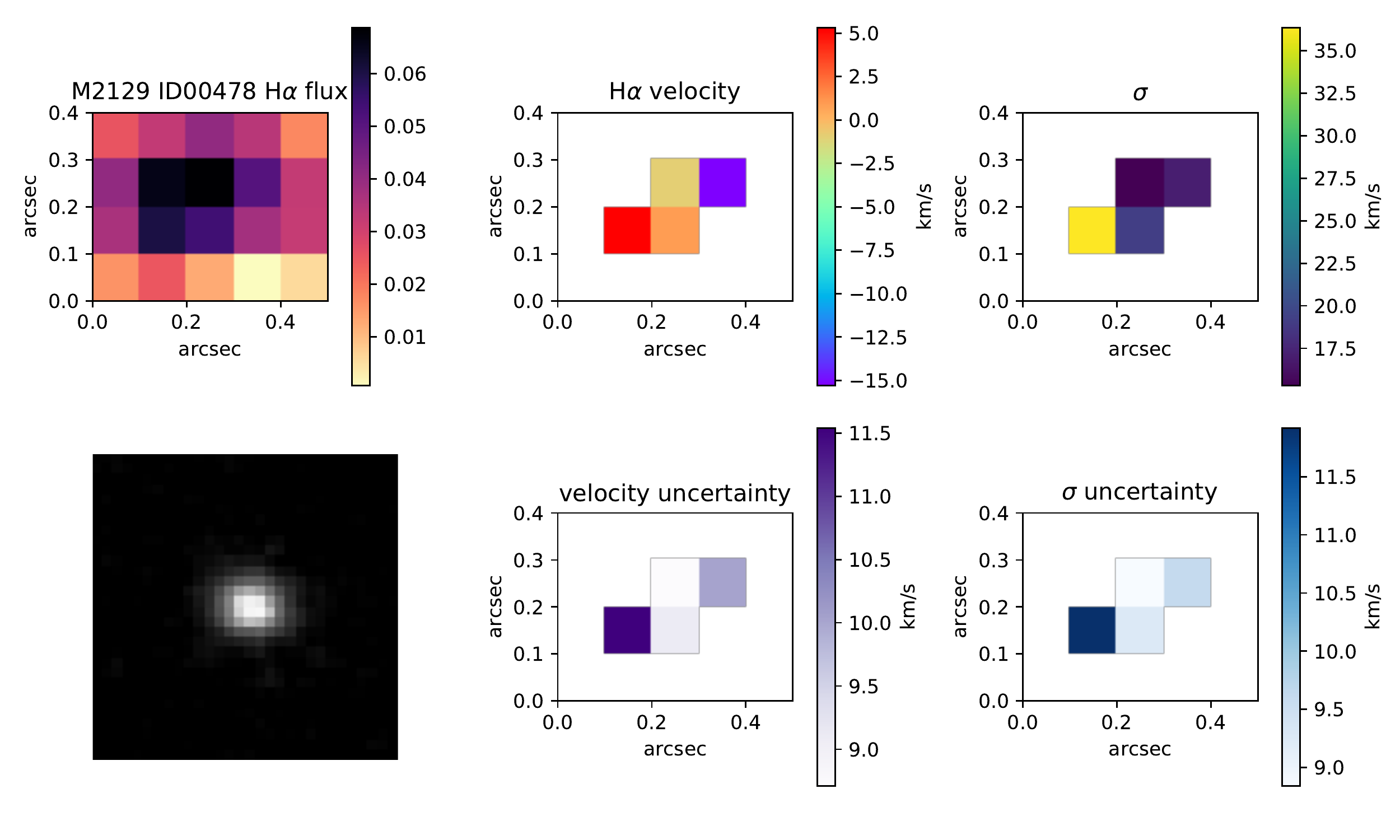}
  \caption{Same as Figure \ref{fig:appendix}, for M2129 ID00478}
\end{figure*}

\begin{figure*}
\centering
  \includegraphics[width=0.9\linewidth]{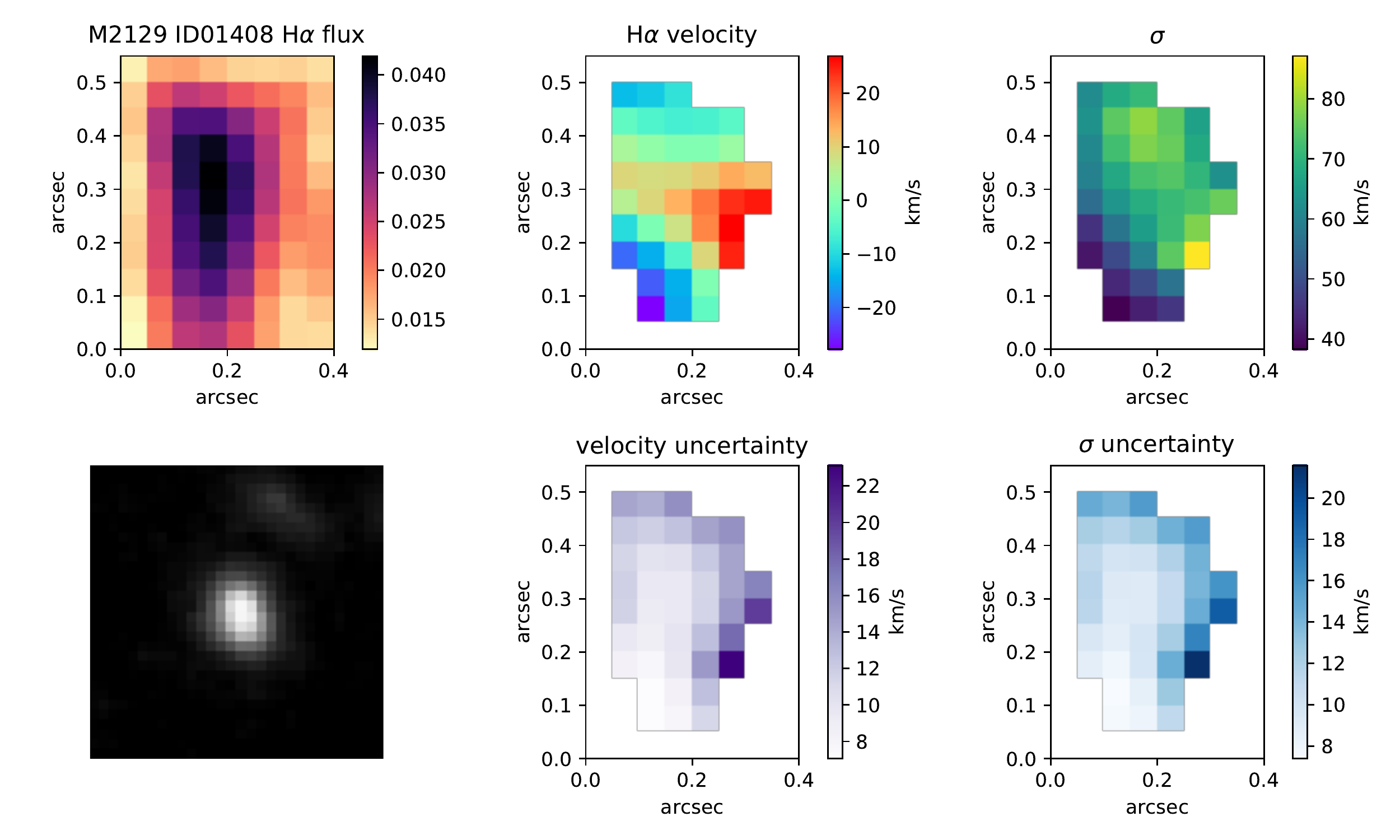}
  \caption{Same as Figure \ref{fig:appendix}, for M2129 ID01408}
\end{figure*}

\begin{figure*}
\centering
  \includegraphics[width=0.9\linewidth]{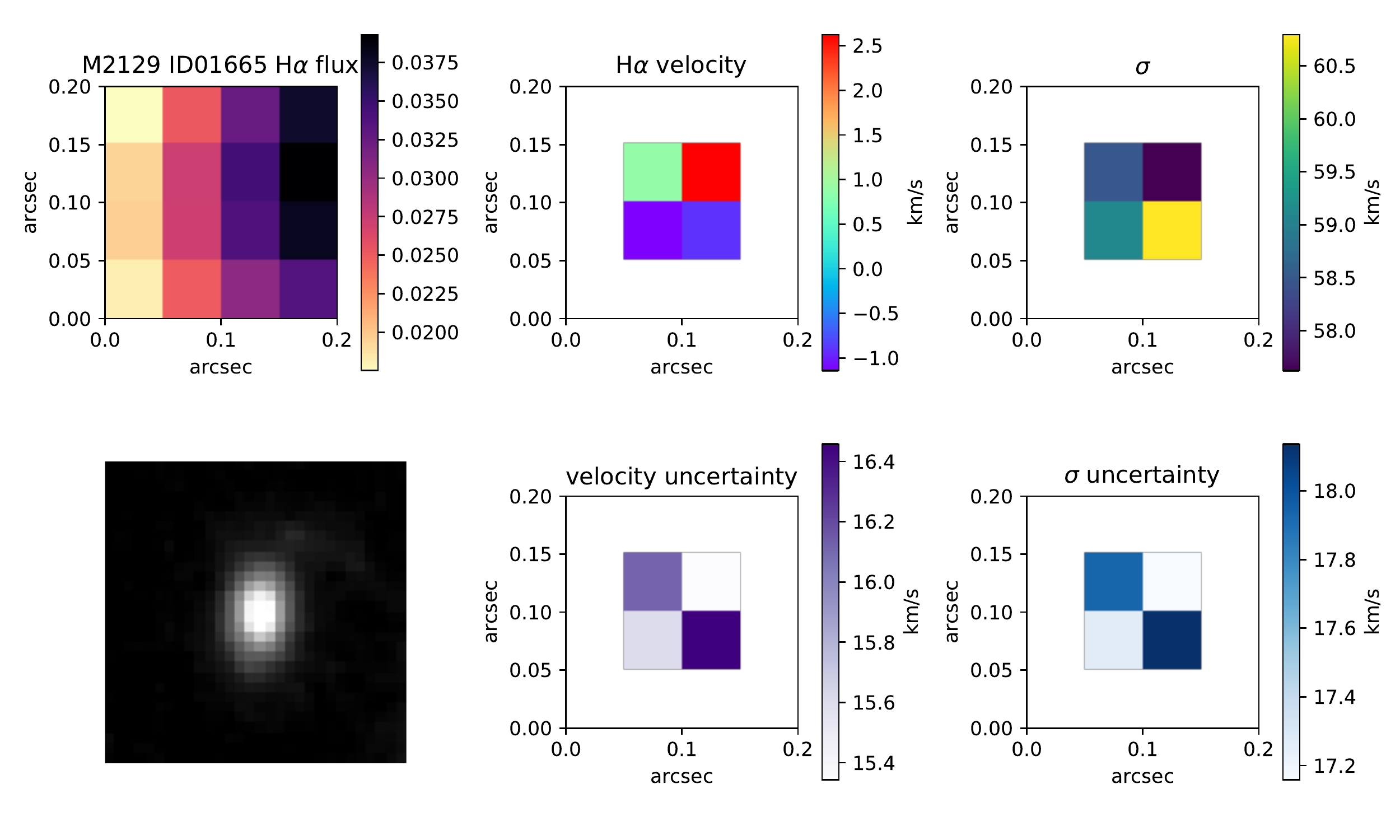}
  \caption{Same as Figure \ref{fig:appendix}, for M2129 ID01665}
\end{figure*}

\end{appendix}

\end{document}